\documentclass[12pt]{cernart}
\usepackage{graphicx}
\usepackage{epsfig}
\usepackage{multirow}
\newcommand{\defmath}[2] {\def#1{\ifmmode{#2}\else\mbox{${#2}$}\fi}} 
\newcommand{\defdecay}[2] {\def#1{\ifmmode{#2}\else{${#2}$}\fi}}
\newcommand{\E}[1] {\mathrm{\times 10^{#1}}}
\newcommand{\amp}[1] {\mathrm{A}({#1})}
\newcommand{\Br}[1] {\mathrm{\Gamma}({#1})}
\newcommand{\expo}[1] {\mathrm{e}^{#1}}

\defdecay{\k} {K}
\defdecay{\kz} {K^0}
\defdecay{\kbar} {\overline{K}^{_{\scriptstyle 0}}}
\defdecay{\kone} {K_1}
\defdecay{\ktwo} {K_2}
\defdecay{\ks} {K_S}
\defdecay{\kl} {K_L}
\defdecay{\ksl} {K_{S,L}}
\defdecay{\KL} {\kl}
\defdecay{\KS} {\ks}

\defdecay{\pipi} {\pi\pi}
\defdecay{\twopi} {2\pi}
\defdecay{\kpipi} {\k \rightarrow \pipi}
\defdecay{\ktwopi} {\ks \rightarrow \twopi}
\defdecay{\kspipi} {\ks \rightarrow \pipi}
\defdecay{\kstwopi} {\kl \rightarrow \twopi}
\defdecay{\klpipi} {\kl \rightarrow \pipi}
\defdecay{\kltwopi} {\kl \rightarrow \twopi}
\defdecay{\pipipi} {\pi\pi\pi}
\defdecay{\threepi} {3\pi}

\defdecay{\pio} {\pi^0}
\defdecay{\twopio} {2 \pi^0}
\defdecay{\piopio} {\pi^0 \pi^0}
\defdecay{\pipin} {\piopio}
\defdecay{\ktwopio} {\k \rightarrow \twopio}
\defdecay{\kpiopio} {\k \rightarrow \piopio}
\defdecay{\kltwopio} {\kl \rightarrow \twopio}
\defdecay{\klpiopio} {\kl \rightarrow \piopio}
\defdecay{\kstwopio} {\ks \rightarrow \twopio}
\defdecay{\kspiopio} {\ks \rightarrow \piopio}

\defdecay{\pip} {\pi^+}
\defdecay{\pim} {\pi^-}
\defdecay{\twopic} {\pi^+ \pi^-}
\defdecay{\pipic} {\twopic}
\defdecay{\ktwopic} {\k \rightarrow \twopic}
\defdecay{\kltwopic} {\kl \rightarrow \twopic}
\defdecay{\kstwopic} {\ks \rightarrow \twopic}

\defdecay{\piod} {\pi_{D}^0}
\defdecay{\piopiod} {\pi^0 \pi_{D}^0}
\defdecay{\eeg} {ee\gamma}
\defdecay{\pioeeg} {\pi^0 \rightarrow \eeg}
\defdecay{\kpiopiod} {\k \rightarrow \piopiod}

\defdecay{\kethree} {\mathrm{K_{e3}}}
\defdecay{\pienu} {\pi e \nu}
\defdecay{\klethree} {\kl \rightarrow \pienu}
\defdecay{\kmuthree} {\mathrm{K_{\mu 3}}}
\defdecay{\pimunu} {\pi \mu \nu}
\defdecay{\klmuthree} {\kl \rightarrow \pimunu}
\defdecay{\threepio} {3\pi^0}
\defdecay{\klthreepio} {\kl \rightarrow \threepio}

\defdecay{\Lam}{\Lambda}
\defdecay{\Lambar}{\bar{\Lambda}}
\defdecay{\etagg}{\eta \rightarrow \gamma \gamma}
\defdecay{\etathreepio} {\eta \rightarrow \threepio}
\defdecay{\piogg} {\pi^0 \rightarrow \gamma \gamma}

\defmath{\dvertex}{d_{vertex}}
\defmath{\mgg}{m_{\gamma\gamma}}
\defmath{\chisq}{\chi^2}
\defmath{\mone}{m_1}
\defmath{\mtwo}{m_2}

\defmath{\pk}{p_K}
\defmath{\pt}{p_T}
\defmath{\ptp}{{p_T}'}
\defmath{\ptpsq}{{p_T}'^2}
\defmath{\mpp}{m_{\pi\pi}}

\defmath{\asl} {\alpha_{SL}}
\defmath{\asloo} {\alpha^{00}_{SL}}
\defmath{\aslpm} {\alpha^{+-}_{SL}}
\defmath{\Dasl} {\Delta \alpha_{SL}}

\defmath{\als} {\alpha_{LS}}
\defmath{\alsoo} {\alpha^{00}_{LS}}
\defmath{\alspm} {\alpha^{+-}_{LS}}
\defmath{\Dals} {\Delta \alpha_{LS}}

\defmath{\btag} {\beta_{tag}}
\defmath{\btagoo} {\beta^{00}_{tag}}
\defmath{\btagpm} {\beta^{+-}_{tag}}
\defmath{\Dbtag} {\Delta \beta_{tag}}

\defmath{\wpm} {W^{+-}}
\defmath{\woo} {W^{00}}
\defmath{\wooo} {W^{000}}
\defmath{\Dw} {\Delta W}

\defmath{\wt}{W(\tau)}
\defmath{\Dp}{\mathrm{D_p}}

\defmath{\etas}{\eta_S}
\defmath{\etal}{\eta_L}
\defmath{\etasl}{\eta_{S,L}}
\defmath{\lamc}{\lambda^{+-}}
\defmath{\lamn}{\lambda^{00}}
\defmath{\DRint}{(\DR)_{\mbox{\scriptsize intensity}}}
\defmath{\DRgeom}{(\DR)_{\mbox{\scriptsize geometry}}}

\defmath{\R}{R}
\defmath{\DR}{\Delta \R}
\defmath{\epp}{\varepsilon^{\prime}}
\defmath{\vep}{\varepsilon}
\defmath{\epe}{\epp/\vep}
\defmath{\Ree}{\mathrm{Re}(\epe)}
\defmath{\epm}{\eta_{+-}}
\defmath{\eoo}{\eta_{00}}

\defmath{\mum}{\mu\mathrm{m}}
\defmath{\mus}{\mu\mathrm{s}}
\defmath{\degrees}{^{\circ}}
\defmath{\taus}{\tau_S}
\defmath{\taul}{\tau_L}

\defmath{\about}{\sim}
\defmath{\eop}{E/p}
\defmath{\Qx} {Q_x}
\defmath{\twotrack}{2track}
\defmath{\etot}{E_{tot}}
\defmath{\mk}{m_K}
\defmath{\stat}{\mbox{stat}}
\defmath{\syst}{\mbox{syst}}

\begin{document}
\begin{titlepage}
\docnum{CERN--EP/2001--067}
\date{20.09.2001}
\title{
 \boldmath A precise measurement of the direct CP
 violation \\ parameter $\Ree$
}
\begin{Authlist}

NA48 Collaboration \\
\begin{center}
A.~Lai,
 D.~Marras \\
{\em Dipartimento di Fisica dell'Universit\`a e Sezione dell'INFN di Cagliari, I-09100 Cagliari, Italy.} \\[0.2cm]
A.~Bevan\footnotemark[1],
 R.S.~Dosanjh,
 T.J.~Gershon\footnotemark[2],
B.~Hay\footnotemark[3],
G.E.~Kalmus,
 C.~Lazzeroni,
 D.J.~Munday,
M.D.~Needham\footnotemark[4],
E.~Olaiya,
 M.A.~Parker,
 T.O.~White,
 S.A.~Wotton \\
{\em Cavendish Laboratory, University of Cambridge, Cambridge, CB3 0HE, U.K.\footnotemark[5].} \\[0.2cm] 
G.~Barr,
 G.~Bocquet,
 A.~Ceccucci,
 T.~Cuhadar-D\"{o}nszelmann,
 D.~Cundy,
 G.~D'Agostini,
 N.~Doble,
V.~Falaleev,
W.~Funk,
 L.~Gatignon,
 A.~Gonidec,
 B.~Gorini,
 G.~Govi,
 P.~Grafstr\"om,
W.~Kubischta,
 A.~Lacourt,
M.~Lenti\footnotemark[6],
S.~Luitz\footnotemark[7],
J.P.~Matheys,
 I.~Mikulec\footnotemark[8],
A.~Norton,
 S.~Palestini,
 B.~Panzer-Steindel,
D.~Schinzel,
G.~Tatishvili\footnotemark[9],
H.~Taureg,
 M.~Velasco,
O.~Vossnack,
 H.~Wahl \\
{\em CERN, CH-1211 Gen\`eve 23, Switzerland.} \\[0.2cm] 
C.~Cheshkov,
A.~Gaponenko\footnotemark[10],
 P.~Hristov,
V.~Kekelidze,
 D.~Madigojine,
N.~Molokanova,
Yu.~Potrebenikov,
 A.~Tkatchev,
 A.~Zinchenko \\
{\em Joint Institute for Nuclear Research, Dubna, Russian    Federation.} \\[0.2cm] 
I.~Knowles,
 V.~Martin,
H.~Parsons,
 R.~Sacco,
 A.~Walker \\
{\em Department of Physics and Astronomy, University of    Edinburgh, Edinburgh,    EH9 3JZ, U.K.\footnotemark[5].} \\[0.2cm] 
M.~Contalbrigo,
 P.~Dalpiaz,
 J.~Duclos,
P.L.~Frabetti\footnotemark[11],
 A.~Gianoli,
 M.~Martini,
 F.~Petrucci,
 M.~Savri\'e,
M.~Scarpa \\
{\em Dipartimento di Fisica dell'Universit\`a e Sezione    dell'INFN di Ferrara, I-44100 Ferrara, Italy.} \\[0.2cm] 
A.~Bizzeti\footnotemark[12],
M.~Calvetti,
 G.~Collazuol,
 G.~Graziani,
 E.~Iacopini,
 F.~Martelli\footnotemark[13],
 M.~Veltri\footnotemark[13] \\
{\em Dipartimento di Fisica dell'Universit\`a e Sezione dell'INFN di Firenze, I-50125 Firenze, Italy.} \\[0.2cm] 
H.G.~Becker,
H.~Bl\"umer,
D.~Coward,
 M.~Eppard,
 H.~Fox,
 A.~Hirstius,
 K.~Holtz,
 A.~Kalter,
 K.~Kleinknecht,
 U.~Koch,
 L.~K\"opke,
 P.~Lopes~da~Silva, 
P.~Marouelli,
 I.~Pellmann,
 A.~Peters,
S.A.~Schmidt,
  V.~Sch\"onharting,
 Y.~Schu\'e,
 R.~Wanke,
 A.~Winhart,
 M.~Wittgen \\
{\em Institut f\"ur Physik, Universit\"at Mainz, D-55099    Mainz, Germany\footnotemark[14].} \\[0.2cm] 
J.C.~Chollet,
 S.~Cr\'ep\'e,
 L.~Fayard,
 L.~Iconomidou-Fayard,
 J.~Ocariz,
 G.~Unal,
 I.~Wingerter-Seez \\
{\em Laboratoire de l'Acc\'el\'erateur Lin\'eaire,  IN2P3-CNRS,Universit\'e de Paris-Sud, 91898 Orsay, France\footnotemark[15].} \\[0.2cm] 
G.~Anzivino,
 P.~Cenci,
 E.~Imbergamo,
 P.~Lubrano,
 A.~Mestvirishvili,
 A.~Nappi,
M.~Pepe,
 M.~Piccini \\
{\em Dipartimento di Fisica dell'Universit\`a e Sezione    dell'INFN di Perugia, I-06100 Perugia, Italy.} \\[0.2cm] 
L.~Bertanza,
P.~Calafiura,
R.~Carosi, 
R.~Casali,
 C.~Cerri,
 M.~Cirilli\footnotemark[16],
F.~Costantini,
 R.~Fantechi,
 S.~Giudici,
 I.~Mannelli, 
V.~Marzulli,
G.~Pierazzini,
 M.~Sozzi \\
{\em Dipartimento di Fisica, Scuola Normale Superiore e SezioneINFN di Pisa, I-56100 Pisa, Italy.} \\[0.2cm]   
J.B.~Cheze,
 J.~Cogan,
 M.~De Beer,
P.~Debu,
F.~Derue,
 A.~Formica,
 R.~Granier de Cassagnac,
E.~Mazzucato,
 B.~Peyaud,
 R.~Turlay,
 B.~Vallage \\
{\em DSM/DAPNIA - CEA Saclay, F-91191 Gif-sur-Yvette, France.} \\[0.2cm] 
I.~Augustin,
M.~Bender,
M.~Holder,
 A.~Maier,
 M.~Ziolkowski \\
{\em Fachbereich Physik, Universit\"at Siegen, D-57068 Siegen, Germany\footnotemark[17].} \\[0.2cm] 
R.~Arcidiacono,
 C.~Biino,
 N.~Cartiglia,
 R.~Guida,
 F.~Marchetto, 
E.~Menichetti,
 N.~Pastrone \\
{\em Dipartimento di Fisica Sperimentale dell'Universit\`a e    Sezione dell'INFN di Torino,  I-10125 Torino, Italy.} \\[0.2cm] 
J.~Nassalski,
 E.~Rondio,
 M.~Szleper,
 W.~Wislicki,
 S.~Wronka \\
{\em Soltan Institute for Nuclear Studies, Laboratory for High    Energy Physics,  PL-00-681 Warsaw, Poland\footnotemark[18].} \\[0.2cm] 
H.~Dibon,
 G.~Fischer,
 M.~Jeitler,
 M.~Markytan,
 G.~Neuhofer,
M.~Pernicka,
 A.~Taurok,
 L.~Widhalm \\
{\em \"Osterreichische Akademie der Wissenschaften, Institut  f\"ur Hochenergiephysik,  A-1050 Wien, Austria\footnotemark[19].} 
\end{center}
\end{Authlist}
Submitted{ to European Physical Journal}
%
\abstract{
The direct CP violation parameter \Ree\ has been measured from the decay
rates of neutral kaons into two pions using the NA48 detector at the CERN
SPS. With 3.3 million \klpiopio\ events 
collected during the 1998 and 1999 running periods, a result of
$\Ree = (15.0 \pm 2.7)\E{-4}$ has been obtained. The result combined
with the published 1997 sample is $\Ree = (15.3 \pm 2.6)\E{-4}$.
} 
\maketitle
\footnotetext[1]{Present address: Oliver Lodge Laboratory, University of
               Liverpool, Liverpool L69 7ZE, U.K.}
\footnotetext[2]{Present address: High Energy Accelerator Research
               Organization (KEK), Tsukuba, Ibaraki, 305-0801, Japan.}
\footnotetext[3]{ Present address: EP Division, CERN, 1211 Gen\`eve 23, Switzerland.}
\footnotetext[4]{ Present address: NIKHEF, PO Box 41882, 1009 DB
  Amsterdam, The Netherlands.}
\footnotetext[5]{ Funded by the U.K.    Particle Physics and Astronomy Research Council.}
\footnotetext[6]{ On leave from Sezione dell'INFN di Firenze, I-50125
  Firenze, Italy.}
\footnotetext[7]{Present address: SLAC, Stanford, CA., 94309, USA.}
\footnotetext[8]{ On leave from \"Osterreichische Akademie der Wissenschaften, Institut  f\"ur Hochenergiephysik,  A-1050 Wien, Austria.}
\footnotetext[9]{ On leave from Joint Institute for Nuclear Research,
  Dubna, 141980, Russian Federation.}
\footnotetext[10]{Present address: University of Alberta, Edmonton Alberta T6G 2J1, Canada.}
\footnotetext[11]{Dipartimento di Fisica e INFN Bologna, viale
  Berti-Pichat 6/2, I-40127 Bologna, Italy.}
\footnotetext[12]{ Dipartimento di Fisica
dell'Universita' di Modena e Reggio Emilia, via G. Campi 213/A
I-41100, Modena, Italy.}
\footnotetext[13]{Instituto di Fisica Universit\'a di Urbino}
\footnotetext[14]{ Funded by the German Federal Minister for    Research and Technology (BMBF) under contract 7MZ18P(4)-TP2.}
\footnotetext[15]{ Funded by Institut National de Physique des
  Particules et de Physique Nucl\'eaire (IN2P3), France}
\footnotetext[16]{Present address: Dipartimento di Fisica
  dell'Universit\'a di Roma ``La Sapienza'' e Sezione INFN di Roma,
  I-00185 Roma, Italy.}
\footnotetext[17]{ Funded by the German Federal Minister for Research and Technology (BMBF) under contract 056SI74.}
\footnotetext[18]{    Supported by the Committee for Scientific Research grants 5P03B10120, 2P03B11719 and SPUB-M/CERN/P03/DZ210/2000 and using computing resources of the Interdisciplinary Center for    Mathematical and    Computational Modelling of the University of Warsaw.} 
\footnotetext[19]{    Funded by the Austrian Ministry of Education,
  Science and Culture under contract GZ 616.360/2-IV GZ
  616.363/2-VIII, and by the Fund for Promotion of Scientific Research
  in Austria (FWF) under contract P08929-PHY.}
\end{titlepage}
\renewcommand{\thefootnote}{\arabic{footnote}}

\section{Introduction}
The violation of CP symmetry
was first reported in 1964 by 
J.H.\ Christenson, J.W.\ Cronin, V.\ Fitch and R.\ Turlay, 
who detected a clean signal of \kltwopic\ decays~\cite{disco}. 
CP conservation implies that the \ks\ and \kl\ particles are pure
CP eigenstates and that \kl\ decays only into CP=$-$1 and \ks\ into
CP=+1 final states. 
The observed signal of the forbidden \kltwopi\ decays (CP=$+$1) indicates that 
CP is not a conserved symmetry.

CP violation can occur via the mixing of CP eigenstates,
called {\em indirect\/} CP violation,
represented by the parameter \vep. 
CP violation can also occur in the decay process itself,
through the interference of final states with different isospins. 
This is represented by the parameter \epp\ 
and is called {\em direct\/} CP violation. 
L.\ Wolfenstein in 1964~\cite{wolf} 
proposed a super-weak force responsible for $\Delta$S=2 transitions, so that
all observed CP violation phenomena come from mixing and \epp =0.
In 1973, Kobayashi and Maskawa proposed a matrix representation 
of the coupling between fermion families~\cite{kob}.  
In the case of three fermion generations, 
both direct and indirect CP violation are naturally accommodated in their model, 
via an irreducible phase. 

The parameters \vep\ and \epp\ are related to the amplitude ratios 
\begin{eqnarray*}
\epm = \frac{ \amp{\kltwopic} }{ \amp{\kstwopic} } = \vep +  \epp   
\end{eqnarray*}
and 
\begin{eqnarray*}
\eoo = \frac{ \amp{\klpiopio} }{ \amp{\kspiopio} } = \vep - 2\epp
\end{eqnarray*}
\hspace{-.18cm}
which represent the strength of the CP violating amplitude 
with respect to the CP conserving one, in each mode. 
Experimentally, it is convenient to measure the double ratio \R\ which is
related  to the ratio \epe:
\begin{equation}
\label{eq:doub}
\R = \frac{ \Br{\klpiopio} }{ \Br{\kspiopio} } / \frac{ \Br{\kltwopic} }{ \Br{\kstwopic} }
  \approx 1 - 6 \times \Ree
\end{equation}
By the mid-1970s, experiments had demonstrated that CP violation in the
neutral kaon system is dominated by mixing, with the 
limit $\Ree \le 10^{-2}$\cite{pioneer}. On the other hand, theoretical 
work showed that 
direct CP violation in the Standard Model could be large enough to be 
measurable \cite{firstthe}. This stimulated experimental
effort with sophisticated detectors to measure $\Ree$.
The first evidence for the existence of a direct component of CP violation 
was published in 1988~\cite{evidence}.
In 1993, two experiments  published their final results without
a conclusive answer on the existence of this component. 
NA31~\cite{na31} measured $\Ree=(23.0\pm6.5)\E{-4}$, 
indicating a $3.5\sigma$ effect.  
The result of E731~\cite{e731}, $\Ree=(7.4\pm5.9)\E{-4}$, 
was compatible with no effect. 
Recently, a new more precise generation of experiments  
announced results from samples of their total statistics. 
The KTeV collaboration measured an effect of $\Ree=(28.0\pm4.1)\E{-4}$
~\cite{res1} and NA48 published a first result of \Ree=$(18.5\pm7.3)\E{-4}$~\cite{res2}. 
These  observations confirmed the existence of a 
direct CP violation component. Current theoretical  predictions 
are in the  range up to \about30$\E{-4}$~\cite{theory}.

This paper reports a measurement of \Ree\ with increased precision, 
performed by the NA48 experiment, 
using data samples recorded in 1998 and 1999. 
The corresponding statistics is seven times larger than
that used for the published 1997 result~\cite{res2}.

\section{The method}
\label{sec:method}

Measuring \Ree\ to a precision of $\sim$$10^{-4}$
requires se\-veral million \kl\ and $\ks \rightarrow\pi\pi$
decays. 
A sufficiently lar\-ge flux of kaons is produced by the
high intensity proton beam from the SPS accelerator. 
Data are accumulated using
a fast and efficient data acquisition system,
including a trigger with high rejection power and 
a large capacity data storage system.

The design of the experiment and the analysis method
focus on making the inevitable systematic biases in the event counting
symmetric between at least two of the four components of the double ratio. 
In this way, most of the important systematic effects cancel to first order,
and only the differences between two components 
need to be considered in detail in the analysis.
This allows the systematic uncertainties to be kept sufficiently low.

In order to exploit the cancellations, 
all four modes are collected
at the same time and from the same decay volume. 
To achieve this, simultaneous \KS\ and \KL\ beams are produced
in two targets situated at different distances from the decay volume.
The intensity ratio is such that the \pipi\ decay rates 
from the two beams are comparable. 
The kaon production angles are tuned to
minimise the diffe\-rence in the \klpipi\ and
\kspipi\ decay spectra over a large range
of kaon energies. 
The beam axes are almost collinear, 
both pointing to the centre of the detector, 
so that the decay products illuminate the detector in a simi\-lar way.  
The similarity is further enhanced
by weighting each \KL\ decay by a function of its proper time,
such that the \KL\ decay distribution becomes almost identical to that of \KS. 
The small remaining differences in beam divergences and
beam geometries are corrected using Monte Carlo simulation. 
The small difference between \KS\ and \KL\ slow beam intensity 
variations are
eliminated by weighting the \KS\ events by the \KL/\KS\ intensity ratio.

As a consequence of simultaneous data collection
all losses associated with the detector, 
trigger and reconstruction efficiencies, 
and with the beam activity, 
cancel to a large extent between \KL\ and \KS.
The ratio of $\pipin$ and $\pipic$ decays in each beam is independent
of the absolute flux.

\KS\ decays are distinguished from \KL\ decays by means of tagging. 
Protons directed to the \KS\ target pass through 
a high-rate tagging station (see section~\ref{sec:tagger}).
\KS\ events can be identified by comparing the registered 
proton time to the event time.
Since this method is used for both the \pipin\ and the \pipic\ samples,
the double ratio depends only on the difference in the
\KS\ misidentification probabilities between the two decay modes,
and not on their absolute values.

Backgrounds affect differently each of the four modes in the double
ratio. High resolution detectors are employed to achieve an efficient
background rejection.
Small remaining impurities due to three body \KL\
decays are carefully subtracted. 

After applying selection criteria,
the four \kpipi\ decay modes are counted in a common 
70 to 170~GeV kaon energy interval. 
In this interval, the $\KS$ and $\KL$ decay spectra are similar 
to within $\pm$15\% (Fig.~\ref{fig:spect}). 
In order to reduce the influence of the residual spectrum 
differences on the result, 
the events are separated into twenty bins of kaon energy, each 5 GeV wide. 
The event 
counts in both the \pipic\ and \pipin\ modes are corrected
for the \about 10\% probability of misassigning a \KL\ decay to the \KS\
beam due to the high proton rate in the tagging station. 
After applying all corrections bin
by bin, the result is obtained by averaging the twenty double ratios.
Remaining uncertainties from the reconstruction, mainly due
to uncertainties in the detector calibration and the geometry, 
are evaluated using several independent methods.

\section{Beams}

\begin{figure*}[htb]
 \begin{center}
\includegraphics[width=0.8\textwidth]{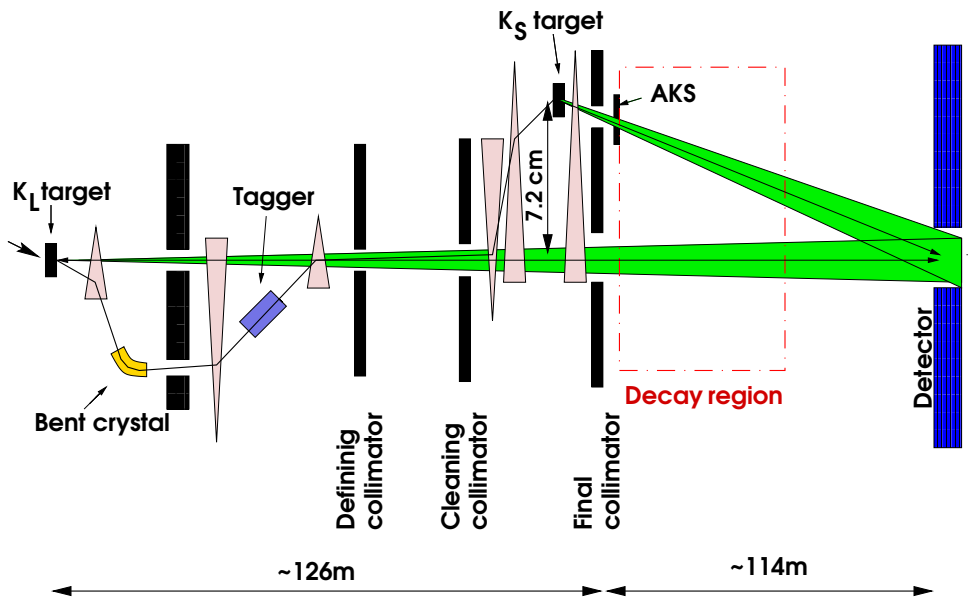} 
 \caption{A schematic view of the beam line (not to scale).}
  \label{beamline}  
 \end{center}
\end{figure*}

The NA48 experiment is installed at CERN and
uses the 450~GeV proton beam delivered by the
SPS. \KL\ and \KS\ beams~\cite{beampaper} 
are produced in  different
targets located 126~m and 6~m upstream of the decay fiducial
region, respectively (Fig.~\ref{beamline}).
\par
The SPS has a cycle time of 14.4~s. 
It is filled with protons in two batches, each 11.5~\mus\ long. 
The beam is accelerated to a momentum of 450~GeV/c 
with a 200~MHz radio-frequency system. 
It is then allowed to debunch, and is extracted 
by means of a slow excitation with 69~\mus\ periodicity 
(3 SPS revolutions) over a spill length of 2.38~s.
The small remnants of the typical frequencies 
(200~MHz, 100~MHz, 87~kHz and harmonics of 50~Hz) 
surviving the filtering and debunching, 
together with burst-to-burst fluctuations of the extraction,  
result in an effective spill length of \about 1.7~s. 
Since the \ks\ and \kl\ beams are produced concurrently, 
the \ks/\kl\ ratio is maintained stable 
throughout the burst to within $\pm10\%$. 
This ensures that both beams are nearly equal 
in their sensitivity to intensity variations of the proton beam.

\subsection{\boldmath The \kl\ beam}
The primary high-flux proton beam ($\about1.5\E{12}$~protons per pulse) 
impinges on a target (a 400~mm long, 2~mm diameter rod of beryllium), 
with an incidence angle of 2.4~mrad relative to the \kl\ beam axis.
The charged component of the outgoing  particles
is swept away by bending magnets. 
The neutral beam passes through three stages of collimation. 
The first ``defining'' collimator, placed 41~m after the target, 
limits the opening angle of the beam.
It is followed, 63 m further downstream, by a second ``cleaning'' collimator, 
which prevents particles scattered or produced on the 
aperture of the defining collimator from reaching the detectors. 
The fiducial region starts at the exit of the ``final'' collimator,
126~m downstream of the target. 
At this point, the neutral beam is dominated by long-lived kaons,
neutrons and photons. 
Only a small fraction of the most energetic of the short-lived 
component (\ks\ and \Lam) survives.

\subsection{\boldmath The \ks\ beam}
The non-interacting protons from the \kl\ target 
are directed onto a mechanically bent mono-crystal of silicon~\cite{crystal}.
A small fraction of protons satisfy the conditions for channelling and
are deflected following the crystalline planes. 
This component passes through a small aperture collimator incorporated in 
a beam dump, which  absorbs the main, undeflected beam of protons.
The fine-tuning of the beam focusing and of the crystal position and angle 
with respect to the incoming
proton direction allows to  select the flux of protons transmitted 
 to be \about3$\E{7}$ per pulse. Use of the crystal enables a 
deflection of 9.6~mrad to be obtained in only 6~cm length, 
corresponding to a bending power of 14.4~Tm. It acts only on the protons
and guarantees a sharp emittance for the selected beam, without cancelling
the deflection of the upstream sweeping magnet on other charged particles.
\par
After the beam dump-collimator, the transmitted protons pass through 
the tagging station (see section~\ref{sec:tagger}) 
which precisely registers their time of passage.
They  are then deflected back onto the \kl\ beam axis, transported through
a series of quadrupoles and finally directed to the \ks\ target (same size as
 \kl) located 72~mm above the \kl\ beam axis. A combination of 
collimator and sweeping magnet defines a neutral beam at 4.2~mrad to the 
incoming protons. The decay spectrum of kaons at the exit of the collimator is 
similar to that in the
\kl\ beam, with an average energy of 110~GeV (Fig.~\ref{fig:spect}). 
\par
The fiducial region begins 6~m downstream of the \ks\ target, 
such that decays are dominated by short lived particles.  At this point, 
the \ks\ and \kl\ beams emerge from the aperture of the final collimators into the common decay region. The whole \ks\ target and collimator system 
is aligned along an axis pointing to the centre of the
detector 120~m away, such that the two beams intersect at this point with 
an angle of 0.6~mrad. The whole decay region is contained in a 90~m long
evacuated tank with a residual pressure of \about 5$\E{-5}$~mbar, 
terminated by a thin composite
polyamide (Kevlar) window of 3$\E{-3}X_{0}$ thickness.

\subsection{Particle rates in the decay region}
It is important to keep the neutral kaon beams as free as possible 
from the contamination of particles, such as muons and scattered neutrons and
photons, produced in the target or along the beam line. 
The 2.4~mrad production angle is chosen to reduce
the neutron flux per useful \kl\ decay. Muons are reduced by the use 
of the bent crystal  and by a further order of magnitude by the 
subsequent sweeping magnets. Beam halo particles are suppressed by means 
of multiple collimation and veto counters. 
Only $2\E{5}$ muons cross the detector for \about$10^{6}$ 
\kl\ decays per spill, while the high fluxes of neutrons (\about$10^{8}$)
 and photons (\about $10^{9}$) in
the neutral beam remain inside the beam pipe such that they do
 not reach the main detector. Since the fiducial decay region is enclosed 
in an evacuated tank
interactions with air are minimised.  

\section{Detectors}

\begin{figure*}[htb]
\begin{center}
\mbox{\epsfig{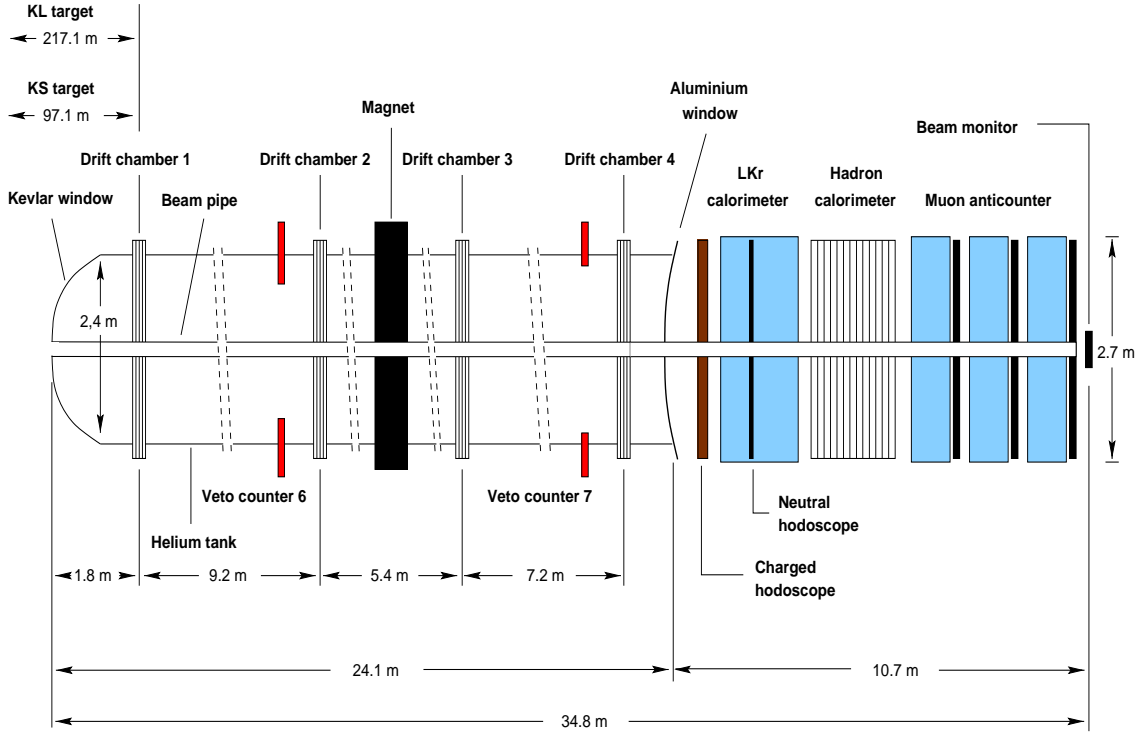}}
\end{center}
\rm
\caption{The detector.}
\label{na48}
\end{figure*}

The detector is shown in Fig.~\ref{na48}~\cite{techpap}. 
Seven annular counters (AKL) surround the fiducial region to 
record photons escaping the acceptance of the main detector.  
Charged particles from decays
are measured by a magnetic spectrometer. This is followed by
a scintillator hodoscope which contributes to
the trigger decision and also gives the precise time of charged decays.
A liquid Krypton
calorimeter is used to trigger and reconstruct K$\rightarrow 2\pi^{0}$ decays. 
It is also used, together with a subsequent iron-scintillator calorimeter
to measure the total visible energy for triggering purposes.
Finally, at the end of the beam 
line, a series of muon counters are used to identify \kl$\rightarrow \pi\mu\nu$ (K$_{\mu3}$) decays. \\
Two beam counters are used to measure the intensity of
the beams. One is located at the extreme end of the \kl\ beam line (\kl\ monitor) and the other (\ks\ monitor) is at
 the \ks\ target station.

\subsection{The tagging station}
\label{sec:tagger}
\begin{figure}[h!]
\begin{center}
\mbox{\epsfig{file=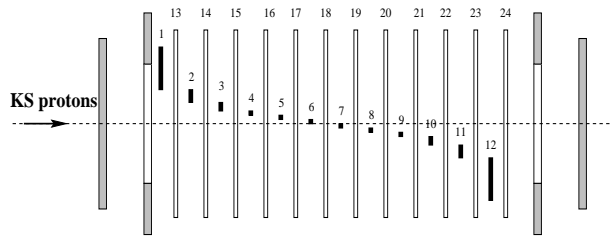,height=8cm,width=3cm,angle=-90}}
\end{center}
\rm
\caption{The arrangement of the counters of the tagging station. Counters 1-12
are oriented horizontally and counters 13-24 vertically.}
\label{taglad}
\end{figure}

The tagging station (or Tagger) is located on the \ks\ proton path after 
the bent crystal. It consists of two scintillator ladders, 
crossing the beam horizontally and vertically~\cite{tagger}. 
All scintillators are 4~mm thick in the beam direction and 15~mm long. 
To equalise the proton counting rate in all channels 
each ladder comprises 12 scintillators of variable widths, 
from 0.2~mm at the centre to 3.0~mm at the edges. 
An overlap of 50~\mum\ between two successive counters
ensures that the beam is completely covered. The scintillators of 
vertical and horizontal ladders
alternate along the beam direction (Fig.~\ref{taglad}). 
A proton crosses at least two of them, one 
horizontal and one vertical. Photomultiplier pulses  are digitised  by an 
8-bit 960~MHz flash ADC module~\cite{tagread}. 
A \about 100~ns window is read out around the trigger time. 
The reconstructed time per counter has a resolution of \about 140~ps,
and two close pulses can be resolved down to 4--5~ns.
Proton times are reconstructed offline by
combining the information from horizontal and vertical counters. 
The coincidence of this time with the event time 
assigns the decay to the \ks\ beam. A demonstration of the
principle is given in Fig. 4 for \pipic\ decays, where the difference
of the event time from the closest proton time  as a function of the 
vertex position in the vertical plane clearly shows the \ks\ and \kl\
beam assignment.
\begin{figure}[h!]
\begin{center}
\vspace{-.65cm}
\hspace{-1.0cm}
\mbox{\epsfig{file=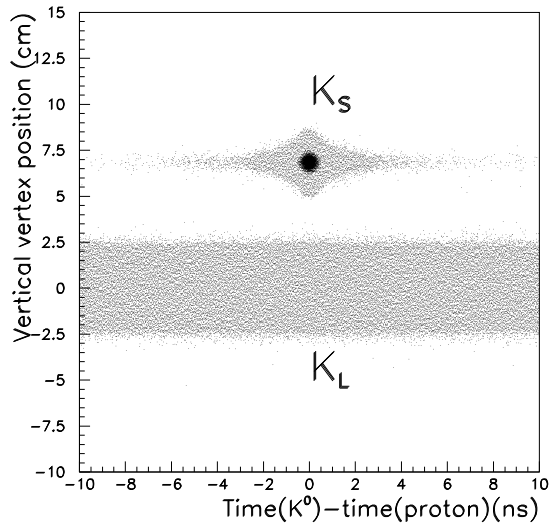,height=8cm,width=8cm}}
 \end{center}
\rm
\hspace{-4cm}
\vspace{-.5cm}
\caption{Time difference between event and its closest proton for $\ks,\kl\rightarrow\pipic$ 
events as a function of the vertical position of their decay vertex.}
\label{tagprinc}
\end{figure}

\subsection{The magnetic spectrometer}
The spectrometer is housed in a tank filled with helium gas at atmospheric pressure. A thin evacuated beam tube 
allows the neutral beam to continue in vacuum.
Two drift chambers (DCH1 and DCH2) are located before, 
and two (DCH3 and DCH4) after, the central dipole magnet. 
These chambers and their interconnecting beam tube are aligned 
along the bisector between the converging \ks\ and \kl\ beam axes.
\par
The integral of the magnetic field is 0.883~Tm, corresponding
to an induced transverse momentum kick of 265~MeV/c in the horizontal plane. 
All components of the field have been measured. 
During the run, the current in the magnet is 
recorded and any relative variation larger than $5\E{-4}$ is corrected for. 
The momentum scale is set adjusting the reconstructed invariant
mass in \pipic\ decays to the nominal $K^{0}$ mass. 
\par
The drift chambers have an octagonal shape 
and an area of 4.5~m$^{2}$~\cite{chambers}. 
Each is made up of four sets of two staggered  
sense wires planes oriented along  four directions,
each one rotated by 45\degrees\ with respect to the previous one, 
and all orthogonal to the beam. 
This permits charged tracks to be reconstructed without ambiguities,
and minimises the effect of wire inefficiencies by providing redundant information. 
The wire material of the eight planes and the filling gas correspond to 
0.4\% $X_{0}$ per chamber. Only four planes of DCH3 are instrumented.
\par
\begin{figure}[h!]
\begin{center}
\vspace{-1.1cm}
\mbox{\epsfig{file=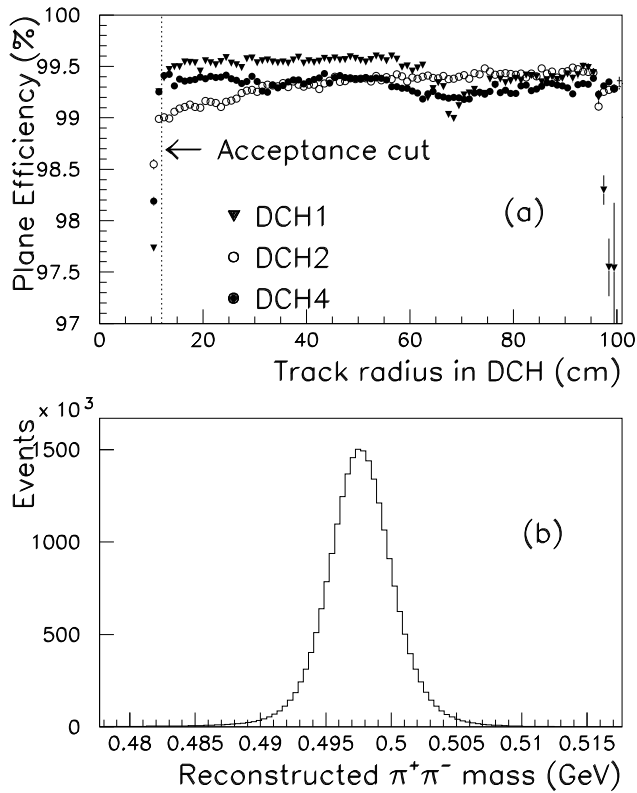,height=12cm,width=12.0cm}}
\end{center}
\rm
\vspace{-.5cm}
\caption{ Plane efficiencies for DCH1,DCH2 
and DCH4 as a function of the track radius (a). Reconstructed \pipic mass (b).}
\label{spectper}
\end{figure}

With such large chambers, 
it is important to know the linear dimensions accurately 
and to have good control of the uniformity. 
The geometric accuracy due to the cumulative uncertainty on the
wire positions is  better than 0.1~mm/m. 
In addition, the  average plane efficiency is measured to be greater than 
99\%, radially uniform to $\pm$ 0.2\% (Fig. \ref{spectper}.a).  

A short drift distance (5~mm, corresponding to a drift time of 100~ns) 
contributes to the high rate capability of this detector. 
Track positions are reconstructed with a precision of 100~\mum\ per view. 
The momentum resolution, obtained from the analysis of special electron runs, 
is $\sigma(p)/p = 0.48\% \oplus 0.009 \times p \%$, where p is in 
$\mathrm{GeV/c}$. The resolution of the invariant \pipic\ mass 
is 2.5~MeV/c$^{2}$ 
(Fig. \ref{spectper}.b). 
The track reconstruction provides an event time with a precision of 0.7~ns. 
This is used to cross-check the main charged event time 
determined by the scintillator hodoscope.

\subsection{The scintillator hodoscope}
A  scintillator hodoscope is placed downstream of the helium tank. 
It is composed of two planes 
segmented in horizontal and vertical strips 
and arranged in four quadrants.
Fast logic combines the quadrant signals, 
which are used in the first level of the trigger for charged events. 
Offline, a combination of hits located at the extrapolated \pipic\ track positions
is used to reconstruct the charged event time with a precision of \about 
150~ps.

\subsection{The liquid Krypton calorimeter}

The liquid Krypton calorimeter (LKr) is a quasi-homoge\-neous detector 
with an active volume of \about 10~m$^{3}$ of liquid Krypton. Cu-Be-Co ribbons
of $40~\mum \times 18~\mathrm{mm} \times 125~\mathrm{cm}$ define 
\about 13000 cells, in a structure of longitudinal projective towers 
pointing to the centre of the decay region~\cite{caloref}. 
The calorimeter is 27~$X_{0}$ long
and fully contains electro-magnetic showers with energies up to 100 GeV. 
The cross section of a cell is about 2~cm $\times$ 2~cm, 
and consists of a central anode in between two cathodes. 
The electrodes are guided longitudinally 
through precisely machined holes in five spacer plates
located every 21~cm. They follow a $\pm$48~mrad zig-zag 
in order to maintain the mechanical stability and to
decrease the sensitivity of the energy resolution to the impact position. 
Good energy response is further guaranteed 
by the initial current readout technique which also provides
a high rate capability. The signals are shaped to \about 75~ns FWHM and 
 are digitised asynchronously by a 40~MHz flash ADC~\cite{caloread}.
The dynamic range of the digitisers is increased by gain-switching 
amplifiers which change the amplification factor 
depending on the pulse height. The calorimetric information readout
is restricted by a zero-suppressing hard-wired programmable algorithm 
to channels belonging to an energy dependent halo around the most 
energetic impact cells.
The calorimeter was operated in a stable way at a high voltage of 
3~kV. Around 0.3\% of the cells were defective and are excluded from the
analysis.
\par
The performance of the calorimeter \cite{calor2000} is studied 
using the electrons from the abundant \klethree\ (\kethree) sample 
recorded along with the \Ree\ data taking.
The electrons are used to improve the cell-to-cell uniformity from 
 0.4\% after electronic calibration to 0.15\%. 
To avoid possible correlations between non-linearity and non-uniformity, 
only electrons in the energy range 25--40~GeV were used for this purpose. 
Cell-to-cell calibration factors were checked  
using photons from \pio\ and $\eta$ decays in special runs.
\par
\begin{figure}[h!]
\begin{center}
\vspace{-1cm}
\mbox{\epsfig{file=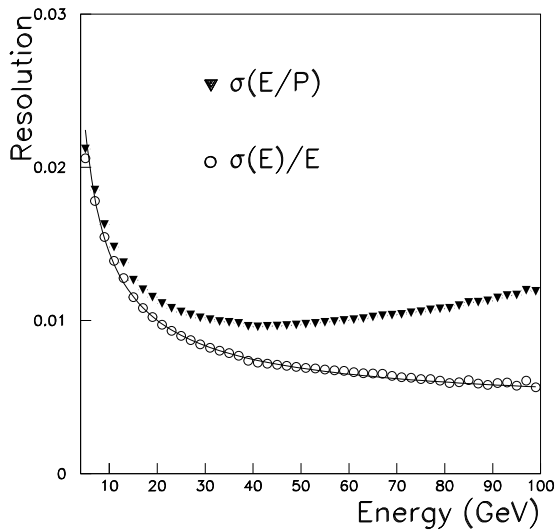,height=8cm,width=8cm}}
\end{center}
\rm\vspace{-.2cm}
\caption{Energy resolution of the LKr calorimeter, obtained with electrons from \klethree\ decays.}
\label{reso}
\end{figure}

Fig.~\ref{reso} shows the resolution of 
the ratio \eop~of energy (from the LKr) 
and momentum (from the spectrometer),  
obtained using \kethree\ events after the inter-calibration procedure. 
Unfolding the measured contribution of the spectrometer 
to the momentum resolution, 
the following energy resolution is obtained, with $E$ in GeV:
\begin{eqnarray*}
\frac{\sigma(E)}{E} =      \frac{(3.2\pm0.2)\%}{\sqrt{E}} 
            \oplus \frac{(9\pm1)\% }{E} 
            \oplus       (0.42\pm0.05)\%
\end{eqnarray*}
The $3.2\%/\sqrt{E}$ sampling term 
is dominated by the fluctuations of the shower fraction  outside  the 
cluster radius used in the reconstruction. The 1/E term is given by the 
total noise in the cluster.
The constant term has contributions from the cell-to-cell calibration, 
the residual gap width variations, the response variation with impact 
distance from the electrodes and the pulse reconstruction accuracy. 

\begin{figure}
\begin{center}
\vspace{-1.cm}
\mbox{\epsfig{file=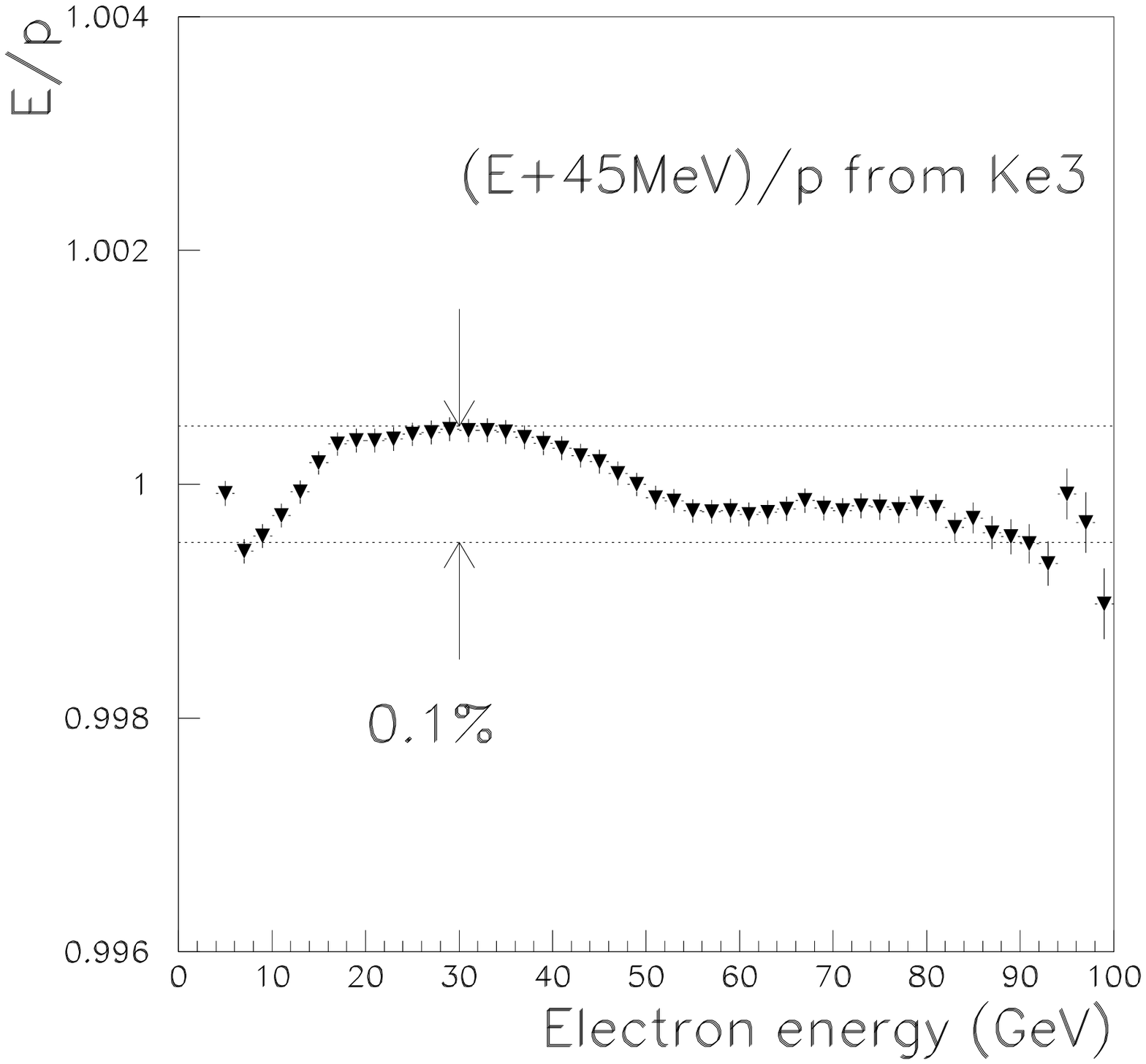,height=8cm,width=8cm}}
\end{center}
\rm
\vspace{-.2cm}
\caption{The linearity of the LKr calorimeter, obtained with electrons from \klethree\ decays.}
\label{linea}
\end{figure}

The energy response is linear to about 0.1\% in the range 5--100~GeV. 
This is shown in Fig.~\ref{linea}, where the average \eop\ is shown for electrons.
To account for losses in the material in front of the calorimeter,  
45~MeV is added to the electron energy as computed by a GEANT-based Monte 
Carlo~\cite{geant}. This simulation also predicts a 0.05\% non-linearity
because of the slight change in the electrode gap width, 
due to the
projective geometry of the calorimeter. This effect is only significant 
for high energy showers which develop deeper inside the calorimeter.
\par
The position resolution of the calorimeter is measured 
using monochromatic electron beams. Comparing the electron centre 
of gravity position from the nine cells around the impact point 
with the extrapolated track point, 
one finds a resolution better than 1~mm in both directions, 
for energies above 25~GeV (the average of the photon spectrum). 
\par
The calorimeter gives an accurate time signal for neutral events,
which is then used together with the tagging station time measurements 
to distinguish \ks\ from \kl. The photon time resolution is of the order 
of 500~ps in the energy range from 3 to 100~GeV.
The \twopio\ event time is known with a precision of \about 220~ps.
Tails coming from misreconstructed times 
have been studied using a single \ks\ beam;  
they are below the level of 10$^{-4}$.
\subsection{The \KS\ anti-counter}
The \KS\ anti-counter (AKS) is located at the exit of the \ks\ collimator.
It is composed of a photon converter followed by three scintillator counters, 
and its main purpose is to veto all upstream decays from the short-lived beam. 
The converter consists of a 2.96~mm thick iridium crystal of 1.8~$X_{0}$~\cite{akspap}, which converts photons into electron pairs keeping the
diffraction probability small. The crystal axis is aligned with the 
beam axis to maximise the pair production.
Given the measured conversion efficiency, only 2.1$\E{-4}$ of the accepted
 \kstwopio\ sample are due to decays occuring upstream of the AKS.
\par
Charged particles, whether from photon conversion in the crystal 
or from \pipic\ decays, are detected by the scintillator counters. 
The time is reconstructed offline with a resolution of about 160~ps.
The inefficiency of the measurement is due to the intrinsic counter 
inefficiency and the dead-time of the subsequent readout system,
which might affect the detection of \kstwopic. 
This results in a total inefficiency of 1.0$\E{-4}$. The final
double ratio is therefore corrected by (1.1$\pm$0.4)$\times 10^{-4}$
for the difference in the AKS inefficiency between \pipin\ and \pipic\ modes.
 
The AKS counter is used  offline to define the beginning of the 
decay region for the \ks\ beam. An event is rejected if a hit is recorded
in time by the second scintillator. The AKS counter
 plays therefore an essential role in the experiment: 
the iridium crystal and the second scintillator give the 
geometrical references needed to control the energy scale and to define
the fiducial region for \ks\ decays into  \pipin\ and \pipic\ respectively.
 
\subsection{Hadron calorimeter and muon counters}
A calorimeter made of iron and scintillators, 
6.7 nuclear interaction lengths thick, 
measures hadronic energy. 
It contributes, together with the LKr calorimeter,
to the total energy trigger. 
Following the hadron calorimeter are the muon counters. 
Three planes of scintillators shielded by 80 cm thick iron walls 
provide timing information that is used offline 
to identify the background from \klmuthree\ (\kmuthree) decays.

\section{Triggers and data acquisition}
\label{sec:tridaq}

The rate of particles reaching the detector is around  500~kHz. The
trigger is designed to reduce this rate, 
with minimal loss from dead time and inefficiencies, 
to several kHz. 
A part of the read-out rate is reserved for
redundant low-bias triggers that collect data used for the direct 
determination of the trigger inefficiencies. 

The trigger decisions are collected in a trigger supervisor system~\cite{tsup}
that records the time of the accepted events, 
relative to the 40 MHz clock signal used to synchronise the experiment 
over its entire 250~m length~\cite{clock}. 
The combined trigger decision is sent back to all the
read-out elements in the form of a time stamp. 
This time stamp arrives
no later than 200~\mus\ after the event occurrence, 
while all data are buffered in the front-end electronics. 
The time stamp is transformed into an address location in the buffer. 
A sufficiently large time interval (100--250~ns depending on the
subdetector) around this location 
is read out and written to tape. 
In case of an occasional pile-up of positive trigger decisions,
consecutive time stamps are queued to allow 
all subsystems to send their data to the event builder before
receiving another time stamp. 
If, nevertheless, one of the subsystems fails to send the complete data,
the whole trigger system is blocked, 
ensuring equal losses for \pipic\ and \pipin\ decays. 
Event building is performed in an on-site PC farm 
and transmitted via a Gigabit network to the CERN computer centre, 
where another PC farm controls the tape writing and carries out
the data reconstruction and the monitoring of the experiment.

\subsection{Trigger for \pipin\ decays}
The trigger for $\pipin$ decays~\cite{nut} operates on the 
analogue sums of signals from $2\times8$ cells of the LKr calorimeter, 
in both horizontal and vertical orientations.
The signals are digitised, 
filtered to reduce noise and summed into 64 columns and 64 rows,
thus providing two projections of the energy deposited.
The summed energy $E$ 
is computed along with
the first and second moments of the energy distribution in each projection,
$M_{1,x}$, $M_{1,y}$, $M_{2,x}$ and $M_{2,y}$. 
The moments are converted into kinematic quantities
using a ``look-up table'' system.
The radial position of the centre of gravity, 
$C=\sqrt{M_{1,x}^2+M_{1,y}^2}/E$, 
is an estimate of the distance between 
the kaon impact point at the calorimeter plane (had it not decayed)
and the beam axis. 
The distance of the decay vertex from the calorimeter, 
$D=E\sqrt{(M_{2,x}+M_{2,y})/E-C^2}/m_K$, 
where \mk\ is the kaon mass, 
is computed to determine the proper time of the kaon decay.
In addition, the number of energy peaks, 
in space and in time, 
in both vertical and horizontal projections, 
is computed in bins of 3 ns. 

The trigger requires
an electro-magnetic energy deposit greater than 50~GeV,
along with $C<15~\mathrm{cm}$ 
and a decay vertex less than 5 \ks\ lifetimes (\taus)
from the beginning of the decay volume.
Another requirement, that there are 
less than 6 peaks within 9 ns in both projections,
helps to reject background from \klthreepio.
This condition is released, however, 
if accidental activity is detected close in time. 

The electronics is implemented in
a pipeline which makes the trigger free of dead time.
The resulting rate of this trigger component is 
2~kHz with a latency of 3~\mus. 

The efficiency is measured with events triggered by a
scintillating fibre hodoscope placed at the depth of about 9.5 $X_0$ 
near the shower maximum
in the LKr calorimeter. It is (99.920$\pm$0.009)\%, 
with no significant difference 
between \ks\ and \kl\ decays. 
Therefore no correction to the \Ree\ measurement is applied. 
The inefficiency is dominated 
by losses from an unresolved pile-up with accidental hits 
and by occasional energy mismeasurement.
A cross-check of the efficiency measurement was performed 
with a very loose trigger condition ($E>$15~GeV) 
applied to events collected in single \ks-beam runs.

\subsection{Trigger for \pipic\ decays}

The $\pipic$ decays are triggered with a two-level trigger system. 
At the first level, 
the rate is reduced to 100~kHz by a coincidence of three fast signals:
\begin{enumerate}
\item Opposite quadrant coincidence in the scintillator
  hodoscope (\Qx), where the quadrants are defined
  with some overlap in order to avoid geometric inefficiencies.
  The remaining inefficiency (0.05\%) is due to electronics and 
  scintillator geometry, and is equal
  for \KS\ and \KL\ decays. However, a \Qx\ signal
  cannot be produced in two consecutive 25~ns 
  clock periods, which leads to
  a dead time of 0.5\%. Signals in a sufficiently large time window are 
  recorded to allow the dead time to be applied (offline) 
  on an event-by-event basis also
  to the \pipin\ sample.
\item Hit multiplicity in DCH1 integrated over 200~ns, 
  requiring at least 3 wires hit in at least 3
  views (\twotrack). This has an inefficiency smaller than $10^{-4}$.
\item Total calorimetric energy (\etot), 
  made by summing the electro-magnetic energy
  from the \pipin\ trigger with the hadron calorimeter energy,
  is required to be more than 35~GeV. Owing to the low resolution in the
  time and size of the hadronic energy measurement,
  some good events fail to pass the
  energy threshold or the time coincidence with the \Qx\ and \twotrack\
  signals. The \etot\ efficiency is $(99.542\pm0.018)\%$ for \KL\
  events (proper-time weighted) 
  and $(99.535\pm0.011)\%$ for \KS\ events. The
  correction for the efficiency difference is applied 
  in bins of energy and
  amounts to $(0.9\pm2.2)\E{-4}$ on the average double ratio.
\end{enumerate}

A signal composed of $\Qx \times \twotrack \times \etot + \Qx/D$ 
is sent to the second level trigger with a latency of 5 $\mu$s.
The $\Qx/D$ component, where $D$ denotes a down-scaling factor, 
is added in order to measure the efficiencies of 
the \twotrack\ and \etot\ components. 
An additional down-scaled $\twotrack \times \etot$ trigger signal 
by-passes the second level trigger 
to allow efficiency measurements 
of the $Q_x$ signal and of the second level trigger.

The second level of the \pipic\ trigger~\cite{mbx} consists of 
hardware coordinate builders 
and a farm of asynchronous microprocessors 
that reconstruct tracks using data from DCH1, 2 and 4. 
Triggers are selected if the tracks converge to within 5~cm, 
their opening angle is smaller than 15~mrad, 
the reconstructed proper decay time is smaller than 4.5 \taus,
and the reconstructed $\pi\pi$ mass is larger than 0.95 \mk. 
The latency is variable but does not exceed 100~\mus\ 
and the output rate is 2~kHz. 
The efficiency is (98.319$\pm$0.038)\% for \KL\ (proper-time weighted)
and  (98.353$\pm$0.022)\% for \KS\ decays. 
The inefficiencies are due mainly to DCH wire inefficiencies (1.2\%) 
with a contribution from algorithm imprecision (0.3\%) 
and misreconstructions from accidental hits (0.2\%). 
The correction for the second level trigger efficiency difference 
is applied in bins of 
energy with an average of $(-4.5\pm4.7)\E{-4}$ on \R.

The time available to extract the data 
from the spectrometer read-out ring buffers to the second level trigger 
is limited.
This leads to a 1.1\% dead time. 
The same dead time condition is applied to \pipin\ candidates event by
event, to
ensure that the principle of collecting \pipic\ and \pipin\
decays concurrently is respected.

In order to avoid recording events with high hit multiplicity, an
overflow condition is generated in the drift chamber readout whenever
more than seven hits in a plane are detected within 100 ns. In this
case the front end readout buffers of this plane are cleared and the
time of the overflow is recorded. Both the second level trigger and
the reconstruction of events are affected by this condition.

Overflows are mainly due to showers induced by interactions of
 electrons or photons in the material surrounding the beam pipe in
 the region of the spectrometer and by $\delta$-rays coming from 
 interactions of charged particles with the drift chamber gas. 
 Occasionally they are also generated
 by noisy amplifiers operated with low thresholds. In this case at
 most two neighbouring planes are affected.

  In the offline reconstruction, a window of $\pm$312 ns
 around the event time is required to be free of overflows, both for
 $\pipic$ and  $\pipin$ decays. This time window is
 larger than the sum of maximum drift time and reset time. In the
 $\pipin$ sample 21.5\% of events are removed by the overflow
 condition which reduces the sensitivity to \KS/\KL\ intensity
 variations by an order of magnitude.

\subsection{Other triggers}
\label{sec:othtri}

Several other triggers were collected continuously during data taking
for systematic studies:
\begin{itemize}
\item A trigger for \threepio\ decays, 
  given by the down-scaled \pipin\ trigger
  without the peak condition, used for \ks\ tagging studies.
\item A trigger for \piopiod\ (where $\piod$ stands for the Dalitz
  decay $\piod \rightarrow \eeg$) decays, 
  combining the information from the LKr calorimeter 
  and the spectrometer, used to test \KS\ tagging and the energy
  scale.
\item Beam monitor triggers 
  used to record the accidental activity,
  with rates proportional to \KL\ and \KS\ decay rates. 
  Beam monitor signals are down-scaled and delayed by 69~\mus\, 
  which corresponds to the periodicity of the slow proton extraction 
  (3 SPS revolutions).
\item Calibration triggers used to monitor and calibrate the LKr
  calorimeter, the tagging station and the scintillator hodoscope.
\end{itemize}

\section{Data samples}

The NA48 experiment collected data for the \Ree\
measurement in SPS running periods during the
summer of three consecutive years: 1997--1999. 
The data from the first running period yielded 0.49 million 
\kltwopio\ events. The result was published in
\cite{res2}. 

In the year 1998, the total number of \kltwopio\ events 
collected during 135 days of running was 1.1 million. 
In the year 1999, 
an upgrade of the trigger and the event builder PC farm,
as well as an increase in the operational stability of the detectors
and electronics,
contributed to obtaining smaller dead time and higher data taking
efficiency. 
In addition, the SPS spill length was increased by
10\%. This allowed the experiment to collect  
2.2 million \kltwopio\  events in 128 days. 
In all run periods the polarity of the magnetic field 
was regularly inverted to
allow systematic checks on \pipic\ decay reconstruction.

In the year 2000, a special run took place to
 cross-check the \KS\ tagging systematics. 
In this run, the NA48 detector operated with vacuum
in place of the spectrometer, 
and with the \KS-protons swept away after their passage 
through the tagging station. 
This allowed a direct measurement of the accidental coincidence rate
between the \kltwopio, \threepio\ decays 
and the protons passing through the Tagger. 

Along with simultaneous \KL\ and \KS\ beam runs,
several auxiliary runs were
dedicated to various systematic checks. 
Data with muon or with \KL\ beams without the spectrometer magnetic field 
were taken for alignment purposes. 
Scans with a monochromatic collimated electron beam 
were used for spectrometer and calorimeter calibration and alignment. 
Runs with either \KS\ or \KL\ beam only were taken on a regular basis 
to verify the \KS\ tagging performance 
and to check the data quality in single beam conditions. 
Tests of the LKr calorimeter calibration were carried out 
using data from runs with two thin polyethylene targets
exposed to a \pim\ beam producing \piogg, \etagg\ and 
\etathreepio\ decays at precisely defined vertex positions.

\section{Decay identification}
\label{sec:decide}

The raw data amount to 170~TBytes. 
After decoding and hit reconstruction, 
the data were filtered and compacted in several steps 
to allow for the iterative improvement of 
calibrations, alignment and corrections. 
The last step, before the final event selection, 
was sufficiently fast that it could be repeated several times 
following the refinements of the corrections.

\subsection{\boldmath Reconstruction and selection of \pipin\ events}
\label{sec:recneu}

The reconstruction of \pipin\ events is based entirely on data from
the LKr calorimeter. 
The time and height of the pulses are measured using
a digital filter technique. 
The first calibration is performed using a calibration pulser system. 
The cell response is inter-calibrated comparing 
energy and momentum (\eop) of electrons from \kethree\ data. 
Further checks and fine tuning of the inter-calibration by $\sim$0.1\%
were performed with $\pio$s produced in a \pim\ beam by adjusting 
the reconstructed vertex of the photon
pairs to the target position. 

Small drifts of the pedestal
due to temperature effects are monitored and corrected. 
A pile up of signals within 3~\mus\ causes the pedestal to shift.
These shifts are detected by comparing ADC samples, 
stored in a buffer before the trigger, with the
average pedestal level. 
If there is a significant difference between
the two, then the stored
samples are used; 
otherwise, the average pedestal is taken.
This procedure minimises 
the influence of noise on the pulse height measurement.

Photon showers are found by looking for maxima in the digitised pulses
from individual cells in both space and time, and accumulating the
energy within a radius of 11~cm.
The shower position is derived 
from the centre of gravity
of $3 \times 3$ central cells. 
Both energy and position measurements
are corrected for a dependence on the distance of the impact point to
the electrodes using data from electron beam scans. 
The transverse scale of the calorimeter is checked using
\kethree-electron tracks and residuals of $\sim$200
$\mu$m are applied as a correction to cluster positions.
In order to account for
deviations from the projectivity of the calorimeter, 
the cluster positions are recomputed at the shower maximum depth. 
The expected shower depth is estimated from \kethree\ data 
comparing reconstructed and extrapolated electron shower positions
and extrapolated to photons using
Monte Carlo simulations. 
 
Overlapping showers are separated using expected
sho\-wer shapes from a GEANT Monte Carlo simulation. 
The quality of the shower shapes
is tested with data from electron beam scans. 
Based on these data, 
an additional correction is applied at the reconstructed shower level.
Energy losses at the borders of the calorimeter are also accounted for, 
using information from Monte Carlo simulations and electron beam scans.
Zero-suppression bias in showers with energy smaller than 5~GeV is 
reduced by parametrising the ratio of the energies deposited in
a well defined $7\times7$ cell shower box and the total reconstructed
shower energy, as a function of the photon energy and 
the number of cells read out. 
A small decrease of measured energy due to space charge
accumulation during the spill, in average around $1.5\times 10^{-4}$, 
is corrected using \kethree\ data~\cite{spacha}. 
The average energy loss of photons in the material before the
calorimeter was determined with GEANT Monte Carlo to be 15~MeV.
Shower energies are increased by this amount.
A correction for residual
nonlinearity of the energy measurement 
is derived from the parametrised \eop\ distribution of
\kethree-electrons (Fig.~\ref{linea}).

Correct shower reconstruction is ensured by accepting
only showers with energies between 3 and 100~GeV,
and positions more than 15~cm from the calorimeter centre, 
11~cm from its outside borders and 2~cm from any defective cell. 

Any group of four showers, each reconstructed within 5~ns of their average
time, is examined for the \pipin\ signature. 
A minimum distance of 10~cm between the showers 
is required to resolve correctly the overlapping energy deposits. 
The event time is computed from an energy-weigh\-ted average of
time measurements in individual cells, 
using the two cells with the largest energy deposits in each shower. 
Cells with signals more than 5~ns from the average time 
are excluded from the time computation. 

\begin{figure}[htb]
 \begin{center} 
\vspace{-1cm}
  \includegraphics[width=0.49\textwidth]{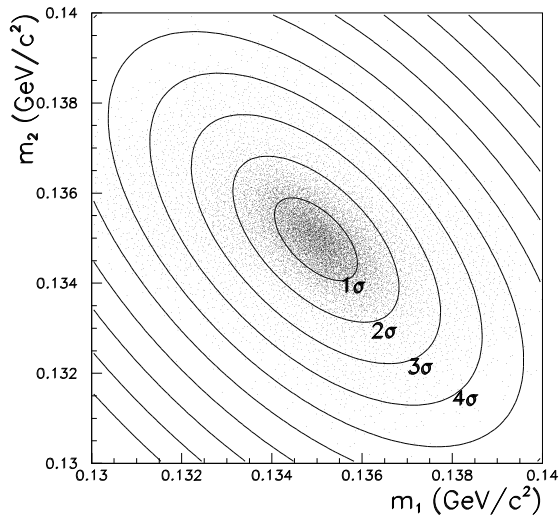}
  \caption{Distribution of $\KS \rightarrow \twopio$ candidates in the
    space of two reconstructed \mgg\ masses,
    \mone\ and \mtwo. The contours correspond to increments by one
    standard deviation.}
  \label{fig:mpi0_2d}   
 \end{center}
\end{figure}

The distance \dvertex\
of the decay vertex from the LKr calorimeter
is computed from the energies $E_i$ and positions $(x_i, y_i)$ of the 
four showers, with the assumption that they come from the decay of a
particle with the kaon mass \mk\ moving along the beam axis:
\begin{equation}
\label{Eq:neuvert}
\dvertex  = \frac{ \sqrt{\sum\limits_i^4\sum\limits_{j>i}^4{E_i}{E_j}
                   \left( (x_i-x_j)^2 + (y_i-y_j)^2 \right) }}{\mk}
\end{equation}
A resolution of $\sim$50 cm is obtained on \dvertex. The vertex
position $z_{vertex}$ is equal to $z_{LKr} - \dvertex$.
The invariant mass of two photons \mgg\ is then computed as
\begin{equation}
\mgg = \frac{\sqrt{{E_1}{E_2}
                   \left( (x_1-x_2)^2 + (y_1-y_2)^2 \right) }}
                   {d_{vertex}}
\end{equation}
The two \mgg\ masses (\mone\ and \mtwo)
are anti-correlated (Fig.~\ref{fig:mpi0_2d}) because \dvertex\ contains
information from all four showers. 
In order to find the best shower pairing, 
a \chisq\ variable is constructed. 
In this variable, \mone\ and \mtwo\ values are
combined in order to remove the anti-correlation.
\begin{equation}
\label{eq:relli}
\chi^2=\left[
        \frac{\frac{(m_1+m_2)}{2}-m_{\pi^0}}{\sigma_+}\right]^2+
        \left[ \frac{\frac{(m_1-m_2)}{2}}{\sigma_-}\right]^2
\end{equation}
where $\sigma_+$ and $\sigma_-$ are the corresponding resolutions,
pa\-ra\-me\-trised as functions of the energy of the least-energetic
photon. 
The average
values of $\sigma_+$ and $\sigma_-$ are 0.42~MeV/c$^2$ and 0.83~MeV/c$^2$, respectively.

\begin{figure}[htb]
 \begin{center}
\vspace{-1cm}
  \includegraphics[width=0.50\textwidth]{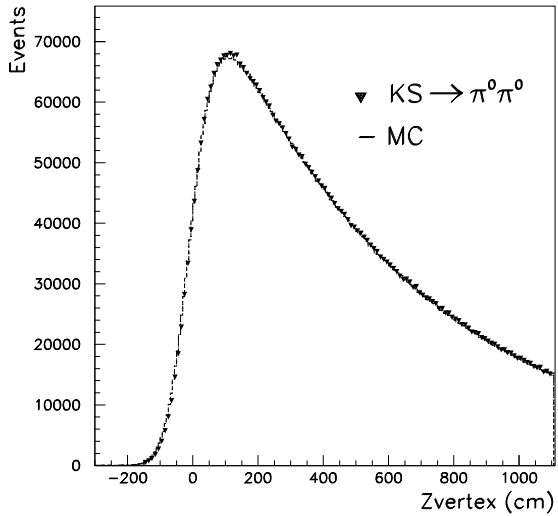}
  \caption{Distribution of the reconstructed vertex position
  $z_{vertex}$ of the
    $\KS\rightarrow \pipin$ candidates at the beginning of the decay
    volume. The origin of the $z_{vertex}$ axis is set to the 
     nominal AKS position.}
  \label{fig:aks_decay_neutral_mc}   
 \end{center}
\end{figure}

As electrons and photons have different energy losses,
the final adjustment of the energy scale is done with photons.
The energy scale is directly coupled to 
the distance scale (Eq.~\ref{Eq:neuvert}).
It is adjusted by comparing the average
vertex position of $\kstwopio$ candidates
at the AKS edge with that produced by the Monte Carlo  
(Fig.~\ref{fig:aks_decay_neutral_mc}), with an accuracy of 3~cm. 
The adjustment of the energy
scale is applied as a function of the run period.
 
The sensitivity of the double ratio to the energy scale is minimised
by the choice of both the kaon energy range and the fiducial decay volume.
At the chosen energy boundaries, 70 to 170 GeV, the  
shape of the decay spectrum favours a cancellation of 
the losses and gains in case of an energy scale shift. 

The total number of selected \kl\ decays is nearly independent of the
distance scale because the acceptance
is only weakly dependent on the longitudinal vertex position.
For \ks\ decays the beginning of the decay volume is 
defined by the AKS counter.
The end of the decay volume is chosen to be 3.5\taus\ 
downstream of the AKS. A 3~cm shift in this cut, 
corresponding to the accuracy of the energy scale adjustment,
together with the effects of energy boundaries and acceptance,
would lead to an error of $2\E{-4}$ in the decay rates.

\begin{figure}[tb]
 \begin{center}
\vspace{-.8cm}
  \includegraphics[width=0.49\textwidth]{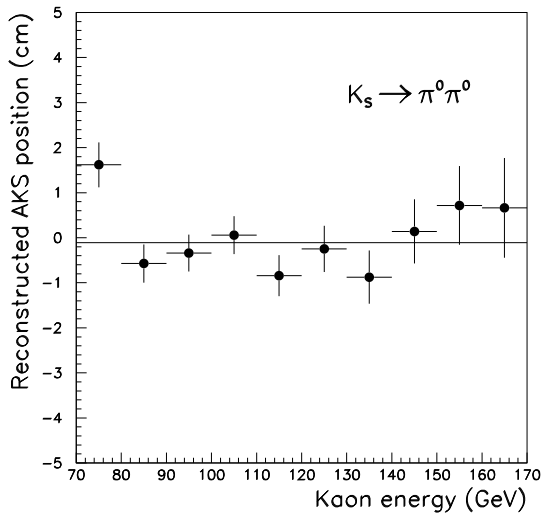}
  \caption{Reconstructed AKS position as a function of kaon energy for
    \kstwopio\ events.}
  \label{fig:aksn_e}   
 \end{center}
\end{figure}

All systematic uncertainties in the \pipin\ reconstruction are
summarised in Tab.~\ref{Tab:neurec}. 
The residual deviation
from linearity in the photon energy
measurement was pa\-ra\-me\-tri\-sed as: 
\begin{equation}
\Delta E=\alpha + \beta E^2 + \gamma rE
\end{equation}
where $E$ is the energy of the photon shower, 
and $r$ is its distance from the centre of the calorimeter. 
An upper limit for each of the coefficients $\alpha$, $\beta$,
$\gamma$ is determined on the basis of the observed nonlinearities in
\kethree\ decays (Fig.~\ref{linea}), in \ktwopio, \threepio\ decays,
and \pio\ and \etagg\ decays from \pim\ beams.
In addition, the stability
of the energy scale is verified as a function of kaon energy
(Fig.~\ref{fig:aksn_e}). 
The sensitivity of \R\ to non-linearities
is determined by modifying the reconstructed
energies in the Monte Carlo. 
The modification takes into account 
also a non-linearity at $E_{\gamma}<6$~GeV observed in \piogg\ decays, 
not covered in the above parametrisation.

The values of the shower depth
used for the position adjustment are
checked by comparing electrons from \kethree\ with Monte Carlo simulations. 
The agreement is better than 2~cm, 
which corresponds to a systematic effect of $1.2\E{-4}$. 
The precision in the sharing of energies in overlapping showers 
is accounted for with a $\pm2.0\E{-4}$ uncertainty on \R\ which is the
size of the correction from the electron beam scan. 
The transverse scale of the LKr calorimeter, 
relative to the spectrometer, 
was verified with \kethree-electron tracks. 
The uncertainty on the double ratio of $\pm2.5\E{-4}$ corresponds to
the difference between these data and measurement during the
calorimeter construction corrected for the termal contraction.

\begin{table}[b]
\begin{center}
\caption{Summary of systematic uncertainties on $R$
in the \pipin\ reconstruction.}
\label{Tab:neurec}
\begin{tabular}{|l|r|}
\hline
& Units of $10^{-4}$ \\ \hline
Energy scale & $\pm$2.0 \\
Nonlinearity: & \\
\hspace{5mm} $|\alpha|<1\times10^{-2}$ GeV & $\pm$2.5 \\
\hspace{5mm} $|\beta|<2\times10^{-5}$ GeV$^{-1}$ & $\pm$2.3 \\
\hspace{5mm} $|\gamma|<1\times10^{-5}$ cm$^{-1}$ & $\pm$1.5 \\
\hspace{5mm} at $E_{\gamma}<6$ GeV & $\pm$1.5 \\
Shower maximum & $\pm$1.2 \\
Energy sharing & $\pm$2.0 \\
Transverse scale & $\pm$2.5 \\
Non-Gaussian tails & $\pm$1.2 \\ \hline
Total & $\pm$5.8 \\ 
\hline
\end{tabular}
\end{center}
\end{table}

Due to hadron photoproduction in the liquid krypton, 
significant asymmetric non-Gaussian tails arise 
in the measurement of energies in about $3\times 10^{-3}$ of
photons. 
These tails are pa\-ra\-me\-trised using electrons and $\pio$ decays
and inserted into the Monte Carlo. 
The uncertainty associated to the effect of non-Gaussian tails is
$\pm1.2\E{-4}$ on the double ratio. 

The scale adjustment is verified by studying
\pio\ and $\eta$ decays from a known origin produced by the \pim\ beam.
Both the \piogg\ and the \etagg\
samples give consistent results. 
For the latter, the $\eta$ mass
is determined from \etathreepio\ decays,
cross-checking the method with  \klthreepio\
decays\footnote{To be published}. 
The linearity of the distance scale along the beam axis is
checked by comparing the results with the two poly\-ethylene targets placed
1462~cm apart. 
The distance measured from data agrees within 1 cm with the nominal
value. This is consistent with the uncertainties on the radial
non-linearity $\gamma$, shower maximum position and energy sharing
from Tab.~\ref{Tab:neurec}.
Another check is performed with \kpiopio\ decays, with subsequent
\pioeeg\ decay,
comparing the vertex reconstructed from the $e^+e^-$ tracks 
to the one derived from the electro-magne\-tic showers. 

\subsection{\boldmath Reconstruction and selection of \pipic\ events}
\label{sec:reccha}

The \pipic\ events are reconstructed from tracks 
using hits in the drift chambers of the spectrometer. 
First, clusters of hits in each pair of staggered planes are found. 
The pattern recognition is based on hit positions 
without using the drift times.
Clusters in DCH1 and DCH2, 
within a range of 10--110~cm from the centre of the chamber, 
are assembled into track segments separately in each coordinate view ($x, y, u, v$). 
The segments are associated with space points in DCH4 with 
the constraint that 
the track is compatible with a straight line
in the $y$ view (vertical direction). 

Each track candidate is converted into one or more tracks 
using the drift time information. 
Drift times are corrected for wire lengths and offsets 
using constants extracted from alignment runs with $\mu$ and \KL\ beams 
without the magnetic field. 
Among tracks sharing the same space point in DCH4, 
the one with the best $\chi^2$ is chosen.
After adding DCH3 information,
the track momenta are calculated 
using a measured magnetic field map and alignment constants. 

A vertex position is calculated
for each pair of tracks with opposite charge
after correcting for a small residual
magnetic field due to the magnetisation of the vacuum tank 
($\sim 2\E{-3}$~Tm). 
The longitudinal vertex position resolution is
about 50~cm, 
whereas the transverse resolution is around 2~mm.
Only tracks with momenta greater than 10~GeV 
and not closer than 12~cm to the centre of each DCH 
are accepted. 
The separation of the two tracks at their closest approach 
is required to be less than 3~cm. 
The track positions, extrapolated downstream, 
are required to be 
within the acceptance of the LKr calorimeter 
and of the muon veto system, 
in order to allow for proper electron and muon identification.

The time of the \pipic\ decay is determined from hits 
in the scintillator hodoscope associated with the tracks.
The events with insufficient information to determine the decay time
accurately are discarded. 
This inefficiency is 0.1\% and is equal for \KS\ and \KL. 

The kaon energy is computed from the opening angle $\theta$ 
of the two tracks before the magnet and from
the ratio of their momenta $p_1$ and $p_2$ assuming a
$\ktwopic$ decay:
\begin{equation}
\label{Eq:eangle}
E_K = \sqrt{\frac{\mathcal{A}}{\theta^2}
        ( m_K^2 - \mathcal{A} m_{\pi}^2 )}
\end{equation} 
where
\begin{equation}
\mathcal{A} = \frac{p_1}{p_2} + \frac{p_2}{p_1} + 2
\end{equation} 
The energy range and the decay volume are defined to be
the same as in the \pipin\ mode.
The small distance between 
the effective positions of
the AKS converter and the AKS counter, 
$(2.2\pm0.1)$~cm is taken into account.

In the \pipic\ mode, 
the distance scale depends only on the
knowledge of the relative geometry of DCH1 and 2.
Likewise, the uncertainty of the energy scale depends on the geometry, 
because the influence of the magnetic field uncertainty cancels in
Eq.~\ref{Eq:eangle}. 
The distance scale in the \pipic\ event reconstruction can be checked,
as in \pipin\ decays, 
by comparing the position of the edge of the  
$\kstwopic$ decay vertex distribution 
with the nominal position of the AKS counter. 
A small mismatch of 2.5~cm is found in the data 
corresponding to a shift in the distance of the two chambers 
of 2~mm or to a 20~\mum\ relative transverse scale mismatch. 
Neither of the two can be excluded. 
A correction of $(2 \pm 2)\E{-4}$, 
corresponding to a 2.5~cm distance scale change, 
is applied to the double ratio.
 
\subsection{\boldmath Background rejection and subtraction in the \pipin\ sample}
\label{sec:neubkg}

The background to the \pipin\ signal comes uniquely from
\klthreepio\ decays.
It is largely suppressed by requiring no
additional showers within $\pm$3~ns around the event time. 
In order to avoid possible losses due to noise,
only showers with reconstructed energy higher than 1.5~GeV are considered for this cut.
This does not affect the suppression of \threepio\ decays since, 
for kinematic reasons, 
very few photons with energy below 1.5~GeV coming from such decays can reach
the LKr calorimeter.
The remaining background consists of events with 
escaping or overlapping photons, 
resulting in only four reconstructed showers. 

\begin{figure}[h!]
\vspace{-1cm}
 \begin{center}
  \includegraphics[width=0.49\textwidth]{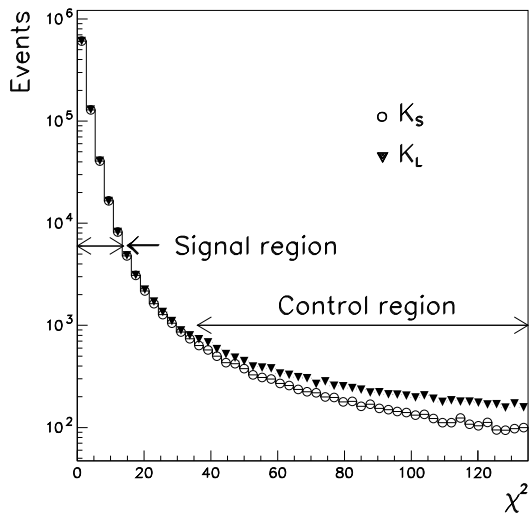}
  \caption{The comparison of the \chisq\ distributions for \KL\ and
     \kstwopio\ candidates, showing the excess due to the
     \threepio\ background in the \KL\ sample.}
  \label{fig:bkn_relli}   
 \end{center}
\end{figure}

Further suppression is achieved by requiring that the \chisq\ 
(Eq.~\ref{eq:relli}) is less than 13.5, 
which corresponds to 3.7$\sigma$ of the \mgg\ resolution.
The loss due to this cut is dominated by photon conversion and is
$\sim$7\%. 
The residual background is evaluated by comparing the properties of the
experimental data in the control region defined by $36<\chisq<135$ 
(Fig.~\ref{fig:bkn_relli}),
for both \KS\ and \KL\ decays, 
with the corresponding Monte Carlo distributions.

In the \KS\ beam, the 4-shower events contain almost exclusively \pipin\ decays,
and the \chisq\ control region is populated by events which have
an energy response or a shower coordinate displacement
outside their normal Gaussian distribution. 
There are essentially two contributing processes. 
The first is a decay with a photon conversion 
(or a \pioeeg\ decay) 
where just 4 showers are reconstructed in the calori\-meter. 
Since no cuts are imposed against
the presence of charged particle tracks, such events can 
end up as \pipin\ candidates. 
Their \chisq\ value can be very large and, 
correlated with it, 
there can be a displacement of the reconstructed kaon decay vertex 
relative to the original longitudinal position.
The second contribution is the occurrence, in about 0.3\% of showers, 
of hadronic photoproduction in their early stage of development
in the liquid Krypton, 
with a consequent reduction of their reconstructed energy. 
Both types of processes, leading to non-Gaussian tails, 
are taken into account in an appropriate Monte Carlo simulation. 
The simulation traces the electron-positron pair through the spectrometer
and, for the effects of hadronic photoproduction, 
makes use of the experimental data obtained 
in calibration runs with electrons. The Monte Carlo for \KS\ gives a good fit 
to the experimental \chisq\ distribution over its full range, 
as well as to the distribution of the longitudinal decay vertex position 
for events in the \chisq\ control region.
On the other hand, 
the experimental \chisq\ distribution for \KL\
clearly exceeds the one predicted by the (exclusively \twopio) Monte Carlo at high
value of \chisq.

The background from \threepio\ decays populates,
to a good approximation,
the \mone-\mtwo\ space evenly, owing to its combinatorial character. 
The \chisq\ is defined
such that each bin corresponds to an equal area 
in the \mone-\mtwo\ plane,
which allows an almost flat extrapolation into the signal region. 
An extrapolation factor of $1.2\pm 0.2$ was
derived from Monte Carlo simulation. 
This simulation, when normalised to the kaon flux, 
agrees with the background level extracted from the data.

The background is subtracted from the \kltwopio\ sample 
in bins of kaon energy.
The resulting correction on the double ratio, 
taking into account all uncertainties, 
is $(-5.9 \pm 2.0)\E{-4}$.   

\subsection{\boldmath Background rejection and subtraction 
in the \pipic\ sample}
\label{sec:chabkg}

\subsubsection{\KS\ sample}

In order to eliminate the background from 
$\Lambda \rightarrow p\pi^-$ 
in the \kstwopic\ sample,
a cut is applied on the track momentum asymmetry, 
\begin{equation}
\frac{|p_+ - p_-|}{p_+ + p_-} < \mathrm{min}(0.62,1.08-0.0052E_K)
\end{equation}
where $E_K$ is the kaon energy in GeV.
This cut is applied to both the \KS\ and the \KL\ samples, 
and its choice is motivated, apart from \Lam\ rejection,
by its improvement of the detector illumination symmetry. 
The residual \Lam\ contamination was verified to be negligible 
by comparing the invariant $m_{p\pi}$ mass distributions 
of \kstwopic\ candidates, 
with $|\mpp-\mk|$ between 3 and 5
sigma of the \mpp\ resolution, 
for opposite signs of the momentum asymmetry.
Due to the large \Lam/\Lambar\ production asymmetry (7/1), in case of
\Lam\ contamination, the two samples would show a significant
difference in the population of this control region which is not
observed.

\subsubsection{\KL\ sample}

In the \kltwopic\ sample, the two semi-leptonic \KL\ 
decay modes, \kethree\ and \kmuthree, 
are the dominant background sources. 
The \kethree\ decays
are suppressed by requiring \eop\ to be less than 0.8. 
\kmuthree\ decays are rejected when
signals in the muon veto system, 
associated with the tracks, 
are found within 4 ns. 
Both cuts are also applied to \KS\ candidates
in order to symmetrise losses of true \pipic\ decays 
with a large electro-magnetic energy deposit (5.0\%) 
or with a $\pi \rightarrow \mu\nu$ decay (1.3\%).

\begin{figure}[htb]
 \begin{center}
  \includegraphics[width=0.49\textwidth]{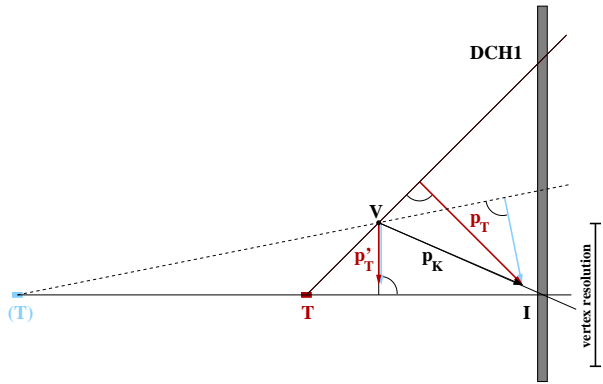}
  \caption{Principle of the transverse momentum calculation viewed 
    in the plane defined by the reconstructed kaon momentum
    \pk\ and the target T; V is the
    decay vertex and I is the reconstructed kaon impact point at
    DCH1. For a vertex misplaced due to
    resolution, \ptp\ remains independent of the target position. }
  \label{fig:pt2_princ}   
 \end{center}
\end{figure}

\begin{figure}[htb]
 \begin{center}
\vspace{-.7cm}
  \includegraphics[width=0.49\textwidth]{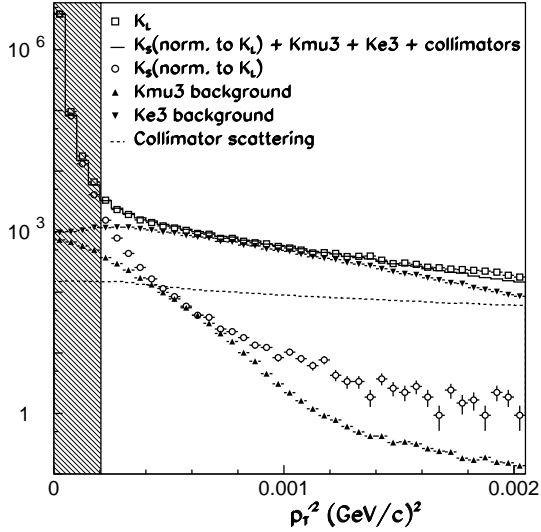}
  \caption{Comparison of the \ptpsq\ tail of the $\KL \rightarrow
    \pipic$ candidates with the sum of all known components.}
  \label{fig:pt-0.002}   
 \end{center}
\end{figure}

Further elimination of the semi-leptonic channels is achieved 
by requiring the invariant mass \mpp\ to be compatible 
with the kaon mass to within 3$\sigma$ of the resolution, 
where the resolution is a function of the kaon energy. 
The cut on \mpp\ also rejects all direct emission $\pi\pi\gamma$ 
decays~\cite{pipig}, present only in the \KL\ beam.
Most of the inner Bremsstrahlung decays are accepted; 
they are, however, \KS\ and \KL\ symmetric. 
This symmetry was tested by Monte Carlo simulation 
to hold to better than $10^{-4}$.

The missing momentum carried away by the undetected neutrino
in the semi-leptonic decays
is reflected in \pt.
The variable \pt\ is the transverse component 
of the reconstructed kaon momentum \pk\
with respect to its flight direction, 
assuming that it comes from the target.
The \ks\ or \kl\ target is chosen according to the 
vertical vertex position.
To equalise the effect of the resolution of the vertex position
reconstruction
for \ks\ and \kl, 
the \ptp\ component of the kaon momentum 
(Fig.~\ref{fig:pt2_princ}) is chosen as a variable for 
background discrimination.
This component is orthogonal to the line TI
joining the target with the reconstructed kaon 
impact point at the DCH1.
The relation between \ptp\ and \pt\ is
\begin{equation}
\ptp = \frac{TV}{TI} \pt
\end{equation}
where TV is the distance from the target to the decay vertex.
By requiring \ptpsq\ to be smaller than 200 MeV$^2$/c$^2$,
most of the semi-leptonic background remaining after the $m_{\pi\pi}$
cut is rejected, 
while \pipic\ event losses are small (0.1\%).
The asymmetry between \ks\ and \kl\ due to non-Gaussian tails is
smaller than $2\E{-4}$. This is included in the reconstruction
uncertainty in Tab.~\ref{tab:syst}. 

In order to subtract the residual \kethree\ and \kmuthree\ background,
two control regions are defined in the \mpp-\ptpsq\ plane.
The first region, 
$9.5<(\mpp-\mk)< 19.0$ MeV/c$^2$ and $300<\ptpsq< 2000$ MeV$^2$/c$^2$, 
is dominated by \kethree\ events, 
while the second, 
$-17.0<(\mpp-\mk)< -12.0$ MeV/c$^2$ and $300<\ptpsq< 500$ MeV$^2$/c$^2$, 
contains roughly equal numbers of \kethree\ and \kmuthree\ events. 
Both regions are chosen such that they contain 
neither $\pi\pi\gamma$ events 
nor collimator scattered events (see section~\ref{sec:colsca}),
and have sufficiently symmetric resolution tails 
in the \KS\ and \KL\ beams.

To model the background
distributions in the control and signal regions, 
a \kethree\ sample is selected with $\eop>0.95$,
and a \kmuthree\ sample is obtained by reversing the muon veto requirement. 
The latter contains also 
$\pi \rightarrow \mu\nu$ decays which are taken into account 
by comparison with a similarly selected \KS\ sample.
The number of true \kltwopic\ decays in
these regions is estimated from the \kstwopic\ sample.
The \kltwopic\ candidates are compared to the model samples 
and the scaling factors that best match the two background model samples are found. 
Their extrapolation into the signal region 
gives a background estimate of $10.1\E{-4}$ for the \kethree\ component, 
and $6.2 \E{-4}$ for \kmuthree. 
The independence of the double ratio on the choice of control 
regions has been tested, 
and all results are compatible within $\pm 2\E{-4}$. 

As a further check, the \ptpsq\ distribution of \kltwopic\ candidates 
is compared over a large \ptpsq\ interval
with the sum of all contributing components (Fig.~\ref{fig:pt-0.002}), 
taking into account 
kaon decays from collimator scattering.
Deviations of around 10\% were shown to come from events with high hit
multiplicity. Their influence on the background estimate is reflected
in an increased uncertainty on the amount of subtracted background.

The background subtraction is applied in bins of kaon energy and the
overall correction on the double ratio is $(16.9 \pm 3.0)\E{-4}$.

\subsection{Collimator scattering correction}
\label{sec:colsca}

Both beams are surrounded by halos of particles from
scattering in the collimators. 
Since the collimators are close to the decay region, 
the scattered particles manifest themselves through \kspipi\ decays. 

\subsubsection{\KS\ beam}

\begin{figure}[htb]
 \begin{center}
\vspace{-1cm}
  \includegraphics[width=0.49\textwidth]{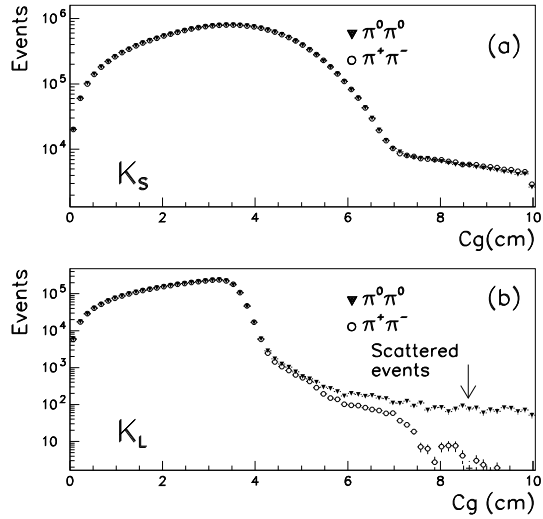}
  \caption{Radial distributions of the centre of gravity for \ks\ and
  \kl\ beams after all cuts.}
  \label{fig:rcog_99}   
 \end{center}
\end{figure}

A beam halo in the \KS\ beam is formed by scattering
in the collimator or in the AKS anti-counter.
It is cut symmetrically in the
\pipin\ and \pipic\ decay modes by requiring the 
radius of the centre of gravity 
(the distance of the virtual kaon impact from the beam axis), 
$C_g = \sqrt{C_{gx}^2 + C_{gy}^2}$,
to be less than 10~cm. 
$C_g$ is defined at the plane of the LKr calorimeter, 
where the two beams cross,
and the cut is applied in all four modes. 
In the \pipin\ mode, $C_g$ is computed from
the positions ($x_i,y_i$) and energies $E_i$ of the four showers. 
\begin{equation}
C_{gx}^{00} = \frac{\sum\limits_1^4{x_i E_i}}{E_K} \hspace{5mm}
C_{gy}^{00} = \frac{\sum\limits_1^4{y_i E_i}}{E_K}
\end{equation}
In the \pipic\ mode, the positions ($x_i,y_i$) are obtained 
by extrapolating the track trajectories before the spectrometer magnet 
to the plane of the LKr calorimeter. The $C_g$ resolutions are almost 
the same for \pipin\ and
\pipic\ decays (Fig.~\ref{fig:rcog_99}a).

The value of the cut is chosen such that only a small part
of the halo ($\sim$1\% of all \KS\ events) 
is rejected. 
The acceptance differences between $\pipic$ and $\pipin$ decays 
are taken into account in the Monte Carlo, 
which contains a parametrised shape of the beam halo. 
The accuracy of the cancellation is verified by varying 
the centre of gravity cut, 
and the uncertainty is taken into account in the acceptance correction.

\subsubsection{\KL\ beam}

The \KL\ beam is defined and cleaned by three collimators.
Double-scattered events
dominate the beam tails. 
Unlike in the \KS\ beam, 
in the \kl\ beam, 
the \ptpsq\ cut rejects more scattered events than the centre of gravity cut.
The \ptpsq\ cut is applied only to \pipic\ decays; 
in \pipin\ decays, the scattered events are removed only by the centre
of gravity cut
(Fig.~\ref{fig:rcog_99}b).

\begin{figure}[h]
\vspace{-1cm}
 \begin{center}
  \includegraphics[width=0.49\textwidth]{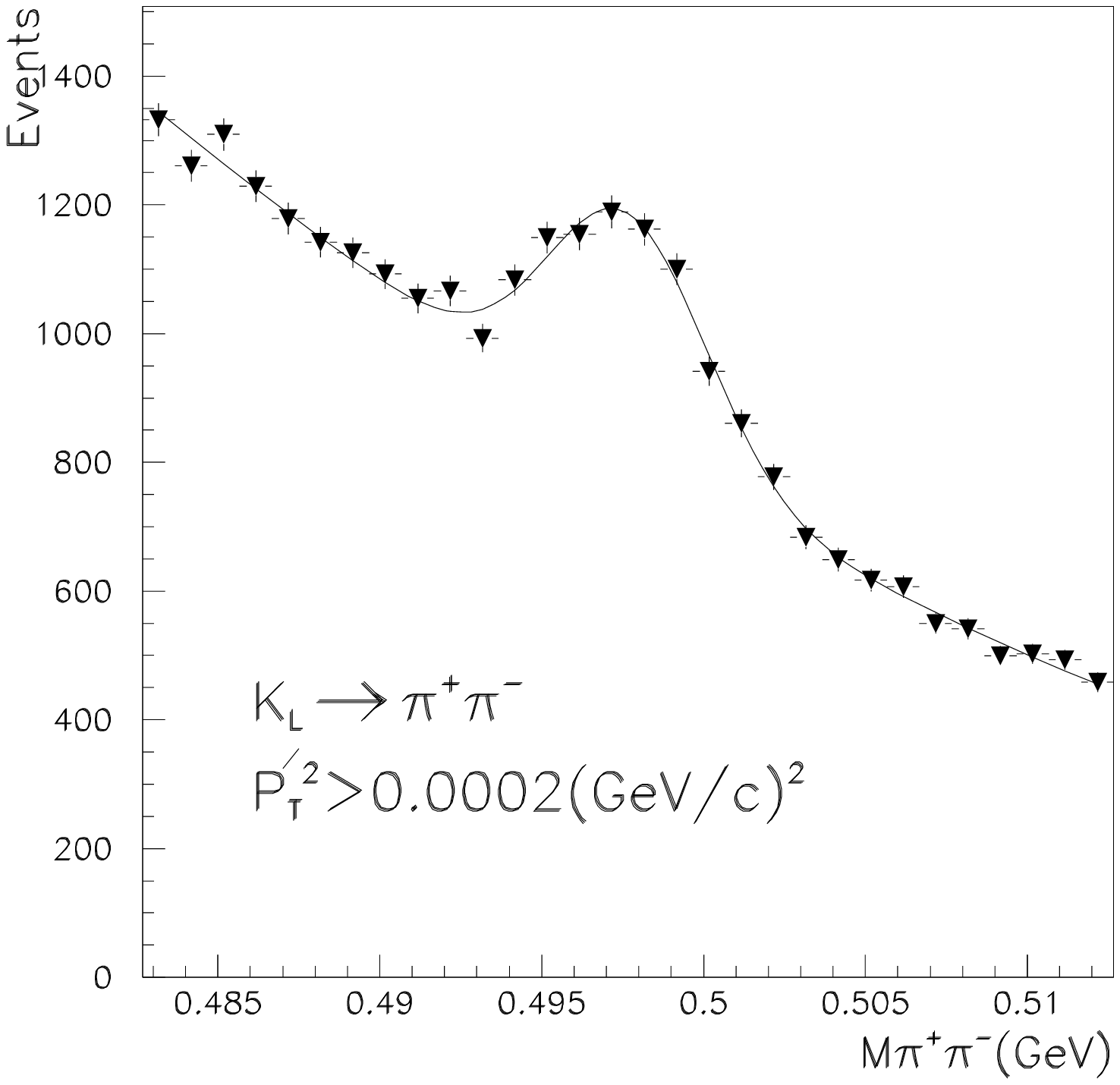}
  \caption{Collimator scattered events at $\ptpsq > 200$ MeV$^2$/c$^2$.}
  \label{fig:mass_hpt}   
 \end{center}
\end{figure}

The correction for this asymmetry is computed from reconstructed 
$\kltwopic$ candidates with an inverted \ptpsq\ cut. 
The scattered events are identified by the \mpp\ invariant mass, 
and the background continuum is subtracted (Fig~\ref{fig:mass_hpt}). 
By extrapolating the computed kaon trajectory
back to the planes of the final or cleaning collimator, 
rings having the collimator radii can be seen.  
This demonstrates the origin of these events. 
Scattered events have a 30\% higher probability of containing an
accompanying shower not associated to the track 
than the decays of kaons coming from the target. 
Since such events are rejected in the \pipin\ selection, 
this factor has to be taken into account in the correction. 
The correction is applied in bins of energy and it amounts to $-(9.6 \pm 2.0)\E{-4}$. 
The error is dominated by the uncertainty on the
accompanying shower factor.

One cross-check of this result involves searching for scattered events 
in the \pipin\ sample with centres of gravity between 5 and 10~cm. 
This region is also populated by \threepio\ background, 
and by decays from kaons scattered in the defining collimator, 
which are confined to low transverse momenta. 
A combination of the three contributions is fitted to the centre of
gravity distribution of \kltwopio\ candidates. 
The model distributions 
of scattered events are obtained from \pipic\ decays, 
and the shape of the 3$\pi^0$ background is obtained 
from events with high \chisq, cross-checked with 5-shower events.
\begin{figure}[h]
 \begin{center}
\vspace{-1.cm}
  \includegraphics[width=0.52\textwidth]{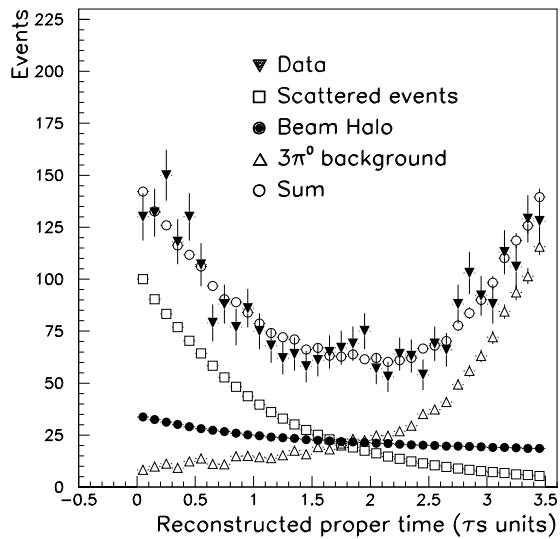}
  \caption{Direct determination of the number of collimator scattered events
    in the $\KL \rightarrow \pipin$ sample with $5 < C_g < 10$ cm 
    from the proper time distribution.}
  \label{fig:rcog00}   
 \end{center}
\end{figure}

A second cross-check 
consists of looking at the proper time distribution of events
with centre of gravity between 5 and 10~cm (Fig.~\ref{fig:rcog00}). 
The scattered events follow the falling \KS\ decay curve, 
whereas the \threepio\ background increases at the end of the decay volume. 
In addition, the distribution of events scattered at the defining collimator, 
which are a mixture of \KL\ and \KS\ decays,
was simulated using the collimator geometries.

All three methods give compatible results within the uncertainty 
of the correction.

\section{\boldmath \ks\ tagging}

\begin{figure}[h!]
\vspace{-1.cm}
\begin{center}
\mbox{\epsfig{file=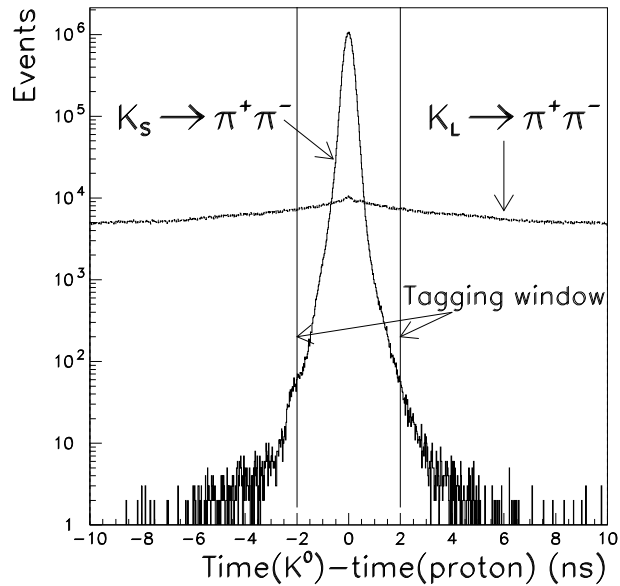,height=9cm,width=9cm}}
\end{center}
\rm
\caption{Time coincidence for charged  \ks\ and  \kl\ events, 
identified by their reconstructed vertex.}
\label{tagdis}
\end{figure}

To associate an event with its parent beam, 
information from the tagging station, 
which records the protons that produce \ks\ particles,
is used.
A decay is labelled \ks\ if a coincidence is found 
between its event time and a proton time measured by the Tagger. 
The good time resolutions inherent to the
detectors guarantee accurate identification. 
Fig.~\ref{tagdis} shows the tagging distributions 
for \ks\ and \kl\ decays to \pipic\, 
which have been identified as such by their vertex position in the vertical 
plane. 
A similar procedure is not possible for neutral decays.  
Tagging is therefore the  only tool available to distinguish \ks\ from \kl.
The selection of \ks\ and \kl\ samples is done by means of tagging for
both the \pipic\ and \pipin\ modes, so that the uncertainties are kept 
symmetric. 
Two kinds of mistagging can occur:
\begin{itemize}
\item A \ks\ decay can be assigned to the \kl\ beam. This
        probability is due to coincidence inefficiencies and 
it is denoted by \asl.
        These inefficiencies are small ($\about 10^{-4}$),
        but they might differ for \pipic\ and \pipin\ decays because the
        respective event times are reconstructed from different detectors.
\item A \kl\ decay can be identified as \ks. 
        This is due to accidental coincidences between the event and proton times, 
        and it is denoted as \als. 
        This probability only depends on the proton rate in the Tagger,
        so, to first order, it affects both \pipic\ and \pipin\ decays equally. 
        It is \about 10\% for the chosen coincidence window.
\end{itemize}
Both mistagging fractions vary according to the width of the coincidence window. 
The tagging window used in this analysis, 
$\pm$2~ns,
was chosen to minimise the total uncertainty on \R\ that comes from mistagging.

\subsection{\boldmath \KS\ tagging inefficiency: \asl}
The tagging inefficiency can be measured accurately in the charged mode, 
because the identification of the decay origin is possible from 
the vertex reconstruction. In identified \ks\ decays, a fraction 
of $(1.63 \pm 0.03)\E{-4}$ 
lies outside the defined $\pm$2~ns coincidence window, as shown 
in the distribution of the difference between the event time provided 
by the hodoscope and the closest proton time (Fig.\ref{tagdis}).
Redundant time information provided by the drift chambers
demonstrates that \about 80\% of these tails are due to Tagger inefficiencies. 
The reconstruction inefficiency of the hodoscope time 
is responsible for the remaining $\about 3\E{-5}$. 
Since mistagging is predominantly due to the Tagger itself,
it affects equally \pipin\ and \pipic\ modes, so that its effect 
cancels in the double ratio.

\begin{figure}[h!]
\begin{center}
\vspace{-1cm}
\mbox{\epsfig{file=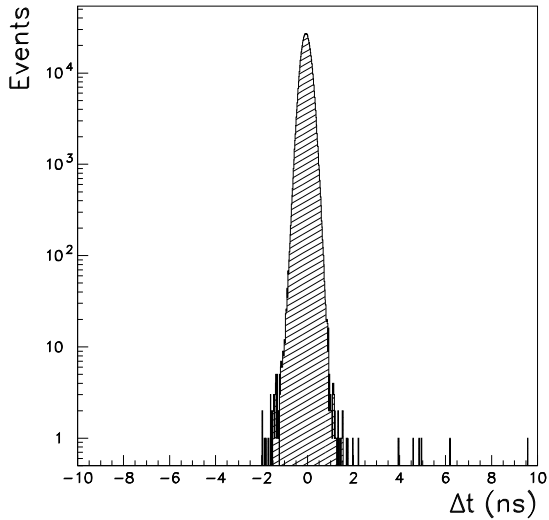,height=8cm,width=8cm}}
\end{center}
\rm

\caption{Difference $\Delta$t between charged and neutral time estimators 
for neutral decays with a conversion.}
\label{convers}
\end{figure}

In order to measure the part of the inefficiency 
associated with the neutral time reconstruction, 
a large sample of \ks\ and \kl\ decays into 2$\pi^{0}$ and 3$\pi^{0}$
is used, where one of the photons converts into an electron-positron pair. 
The usual selection and quality criteria are applied,  
but without requiring the reconstruction of the complete decay. 
Accepted events are allowed to contain between four and seven clusters. 
The two tracks must cross to give a vertex 
and must fulfill the electron identification hypothesis, i.e. $\eop >0.95$. 
The charged time is computed using the standard procedure 
applied to the two electron tracks. 
The neutral time is computed
by the association of central and lateral cell times of all
electro-magnetic showers satisfying some weak criteria 
which ensure that they belong to a single event.  
The event-by-event compatibility  of the charged and neutral time 
is checked within the usual $\pm$2~ns window (Fig. \ref{convers}). 

The number of cases where the two time measurements differ by more than 2~ns 
corresponds to the differential inefficiency 
associated with hodoscope and calorimeter time reconstruction. 
Including  the sensitivity of the neutral time computation to the number 
and to the energy of photons in the systematic uncertainty,
one finds that the tails in the time estimator difference $\Delta$t are
 less than 2.5$\E{-5}$ at a 68\% confidence level.   
The same data sample allows the precise determination of the
absolute value of the neutral tagging inefficiency. 
Convoluting the neutral-charged time difference 
with the \pipic\-$-$Tagger time difference,
the tagging inefficiency for \pipin\ is obtained.  
This inefficiency of (1.64$\pm$0.5)$\E{-4}$, 
has been found to be in good agreement 
with a direct evaluation determined in single \ks\ beam, 
where all events are associated with a proton. 
A third method consists of checking the neutral tagging inefficiency
in an identified \kstwopio\ sample belonging to the main data set. 
For this, \pipin\ data are used with a subsequent \pioeeg\ Dalitz decay.
The vertex position reconstructed from the electron and positron tracks 
allows \ks\ decays to be selected.
All events outside the $\pm$~2ns coincidence window 
correspond to Tagger or calorimeter reconstruction inefficiencies.
The result is in good agreement with the two methods described previously, 
though it remains statistically limited 
because of the small (1.2\%) Dalitz branching ratio. 
Tab.~\ref{asl} summarises the results on \pipin\ tagging inefficiency.
\begin{table}[h!]
\begin{center}
\caption{Different evaluations of \ks~tagging inefficiency for \pipin\ events.}
\label{asl}
\vspace{2mm}
\begin{tabular}{|l|c|c|}
\hline
{\bf{Method}}&{\bf \boldmath {\asloo\ in 10$^{-4}$}} \\
\hline
{\bf{Conversions}}& 1.64$\pm$0.5(stat+syst)  \\
\hline
{\bf \boldmath {Single \ks\ beam}} & 1.90$\pm$0.42(stat) \\
\hline
{\bf \boldmath {\pipin\ with Dalitz}} & 1.9$^{+1.0}_{-0.8}$(stat) \\
\hline
\end{tabular}
\end{center}
\end{table}

The important consideration for the double ratio measurement 
is the mistagging difference between the charged and neutral modes. 
From the previous analyses, 
it can be concluded that there is no measurable effect
within an uncertainty of $\pm0.5\E{-4}$.

\subsection{\boldmath Accidental tagging: \als}

\begin{figure}[h!]
\vspace{-1.cm}
\begin{center}
  \includegraphics[height=13cm]{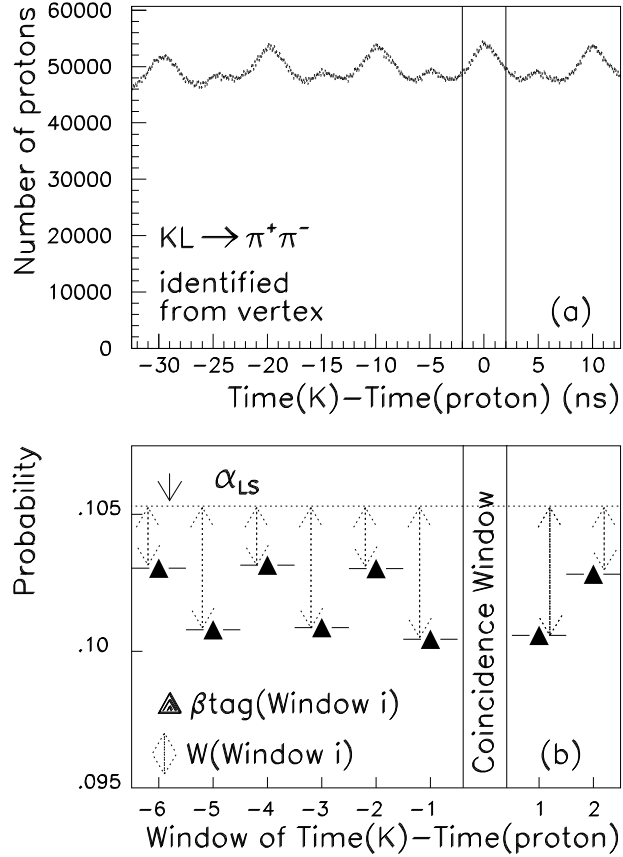}
\end{center}
\rm\caption{(a) The distribution of proton times with respect to
 event time for 
             $K_{L}\rightarrow\pi^{+}\pi^{-}$ identified from vertex. 
         (b) The variables \als, \btag\ and $W$ are schematically shown for 
             $K_{L}\rightarrow\pi\pi$  tagged events.}
\label{wtagp}

\end{figure}

Because of the high rate in the Tagger, 
\kl\ events can have  an accidental coincidence with a proton. 
The probability for this to happen 
is proportional to the width of the tagging window. 
It can be measured directly in charged decays 
by looking at the fraction of those \pipic\ events 
identified from their vertex position as \kl\ decays, 
that also have a proton in the $\pm$2~ns time window (Fig.~\ref{wtagp}a).
On average, it was found that $\alspm = (10.649\pm0.008)\%$.
Such a direct evaluation is not possible for neutral events. 
An indirect method is therefore applied to both the \twopic\ and the \pipin\ modes,
to evaluate the quantity $\Dals \equiv \alsoo - \alspm$,
which leads to a correction to the double ratio.
The \kl\ samples were identified as such by tagging, so
that no proton is detected within $\pm$2~ns of the event time.
With this sample, the quantity \btag\ is computed, 
which is the probability of having a proton within a $\pm$2 ns sideband window,
located before or after the event time (Fig.~\ref{wtagp}b). 
These windows are centred at 5~ns intervals from the event time, 
and so follow the 200~MHz structure of the proton beam. 
Because \btag\ is measured using the fraction of \kl\ events without an 
accidental coincidence in the central window, the proton rate is generally
slightly lower than for the full sample. This is reflected in \btag\
being smaller than \als\  by a quantity $W$. Typically, $W$ is $\sim$10\%
of \als. Mainly due to the 100~MHz beam structure, 
\btag\ varies with the window; 
however, the corresponding $W$ value compensates for this, 
using the relation:
\begin{center}
 $\alsoo-\alspm = (\btagoo + \woo ) - (\btagpm + \wpm)$
\end{center} 
The \btag\ parameters 
are extracted from the same data that are used for 
the double ratio determination. In the \pipic\ mode,
\wpm\ can be computed by comparing \alspm\ and \btagpm\ in all windows, 
using the \kl\ sample identified by vertex position. 
In a similar way, \btagoo\ can be extracted. 
To measure \woo, the assumption is made that
\kltwopio\ and \klthreepio\ decays are tagged identically, 
giving  $\woo=\wooo$. 
The abundant sample of \klthreepio\ decays is then used to derive \woo.
Finally, in order to increase the statistical significance of the measurements,
several 4~ns wide windows are used in the sideband region. 
The results are shown in Tab. \ref{tagpar}:

\begin{table}[h!]
\begin{center}
\caption{Measured values of parameters \Dbtag, \Dw\ and \Dals.}
\label{tagpar}
\vspace{2mm}
\begin{tabular}{|l|c|c|}
\hline
 &{\bf \boldmath {Units of 10$^{-4}$}}\\
\hline
{\bf \boldmath {$\Dbtag = \btagoo - \btagpm $}}         &       $ 3.0 \pm 1.0 (\mathrm{stat})$ \\
\hline
{\bf \boldmath {$\Dw = \woo - \wpm $}}                  &       $ 1.3 \pm 1.1 (\mathrm{stat})$ \\
\hline
{\bf \boldmath {$\Dals = \alsoo - \alspm$}}             &       $ 4.3 \pm 1.4 (\mathrm{stat}) \pm 1.0 (\mathrm{syst}) $\\
\hline
\end{tabular}

\end{center}
\end{table}
The quoted systematic error on \Dals\ takes into account the
small variation of the result
depending on the set of out-of-time windows used in the computation.
Several arguments support the assumption $\woo = \wooo$. 
Firstly, the trigger conditions for \kltwopi\ and \klthreepio\ are similar and 
highly efficient (\about 99.9\%).
Secondly, these two modes are reconstructed by the same detector with similar 
criteria. Finally, in 2000, a special run period was devoted to measure 
simultaneously both $\woo$ and  $\wooo$. After passing through the Tagger, the \ks\ proton
beam was deviated away from the trajectory leading to the \ks\ target, 
so that only particles from the \kl\ beam decayed in the fiducial region. 
All the tagging parameters analysed show good compatibility between
the \twopio\ and \threepio\ modes. In particular it is found:
\begin{eqnarray*}
\woo - \wooo = (-0.8 \pm 2.9)\E{-4}
\end{eqnarray*}
which confirms the initial assumption. Even though there is an overall
agreement of the tagging parameters between the \twopio\ and \threepio\ 
modes, \btagoo, which is the largest part of \alsoo, is directly measured
from the sample  entering in the double ratio, namely \twopio\ events.
The usage of the external \threepio\  sample is restricted to the
determination of the \woo\ parameter only.

\subsubsection{Origins of  $\Dals \ne 0$}
The measured difference  $\Dals=(4.3 \pm 1.8)\E{-4}$  (see Tab.\ref{tagpar})
indicates that the \pipin\ and \pipic\  
samples are recorded in conditions of slightly different intensities, although
the same rejection of trigger dead time and drift chamber overflow is  
applied to both modes.
The reason for this is the higher sensitivity to accidentals in the trigger
conditions and reconstruction of \pipic\ compared to \pipin. The first 
contribution to \Dals, associated with \pipic\ events lost at the trigger,
 is $(1.0 \pm 0.3) \E{-4}$. This has been verified studying events  
collected with fully relaxed trigger. The second contribution, related to the
event losses at the reconstruction, is studied superimposing 
good events from data or simulation to beam monitor triggers, recorded 
in proportion to the intensity
(see section~\ref{sec:overlay}), and it is $(2.5 \pm 0.4)\E{-4}$.

Tab.~\ref{deltaALS} shows the comparison between expected 
and measured \Dals\ values for the available data samples.  
\begin{table}[h!]
\begin{center}
\caption{Values of \Dals\ in the two different
data periods, and comparison with the expectations from known sources.}
\label{deltaALS}
\vspace{2mm}
\begin{tabular}{|c|c|c|}
\hline
\bf Data Sample         & \bf \boldmath Expected \Dals  & \bf \boldmath Measured \Dals \\
                        & \bf \boldmath Units of 10$^{-4}$    & \bf \boldmath Units of 10$^{-4}$ \\
\hline
{\bf{1998}} & 5.1$\pm$1.0 & 8.3$\pm$2.9\\
\hline
{\bf{1999}} & 2.6$\pm$0.6 & 2.6$\pm$2.1\\ 
\hline
{\bf{1998+1999}}& 3.5$\pm$0.5& 4.3$\pm$1.8\\
\hline
\end{tabular}
\end{center}
\end{table}
It is noted that improvements
in the \pipic\ trigger code and the lower instantaneous intensity
allowed by a longer spill, result in a smaller loss for
charged decays and consequently in a smaller \Dals\ in 1999 relative to 1998
data. 
The measured values of \Dals\ and their variation between the two 
years agree well with the expectations within errors. 

\subsection{\boldmath Effect of the \ks\ tagging on \R}
After all cuts, four event samples remain, namely \pipic\ and \pipin\ 
events tagged as \ks, and \pipin\ and \pipic\  events tagged as \kl.
The mistagged fraction of events is subtracted from \ks~and added to \kl. 
The double ratio is then sensitive to \Dasl\ and \Dals: 
\begin{eqnarray*}
\DR \approx -6(\asloo - \aslpm) \hspace{5mm}
\DR \approx 2(\alsoo - \alspm)
\end{eqnarray*}
where the numerical factors depend on the absolute mistagging
probability \als\ and  on the flux ratio \ks/\kl.
The corresponding corrections to the double ratio become:
\begin{eqnarray*}
\DR = (0.0 \pm 3.0)\E{-4}~\mbox{for \Dasl} \\
\DR = (8.3 \pm 3.4)\E{-4}~\mbox{for \Dals} 
\end{eqnarray*}
\section{Acceptance correction} 
\ks\ and \kl\ particles decay with very different lifetimes. 
Despite the use of a common decay region of 3.5\taus,
the decay distributions vary strongly along the beam direction.
The resulting acceptance correction on \R\ can reach about $\pm$10\%,
depending on  the kaon energy. 
To cancel the contribution from the different lifetimes to the acceptance, 
the \kl\ distributions are weighted by a factor \wt, 
where $\tau$ is the proper time of the kaon:
\begin{eqnarray*}
\wt = \frac{ \normalsize I(\tau~\mbox{from \ks\ target}) \rightarrow \twopi }
           { \normalsize I(\tau~\mbox{from \kl\ target}) \rightarrow \twopi }
\end{eqnarray*}
$I(\tau)$ describes the complete kaon decay intensity into \twopi, 
containing \ks, \kl\ and interference components:\\

$I(\tau) = \expo{-\tau/\taus} 
        + |\eta|^2 \expo{-\tau/\taul}\\
~~~~~~~~~~~~~~~~~~~~~~~~~~        + 2 |\eta| \Dp \expo{-\frac{1}{2}\tau(1/\taul + 1/\taus)} \cos(\Delta M\tau-\phi)$ \\

\begin{figure}[h!]
\begin{center}
\vspace{-.8cm}
\mbox{\epsfig{file=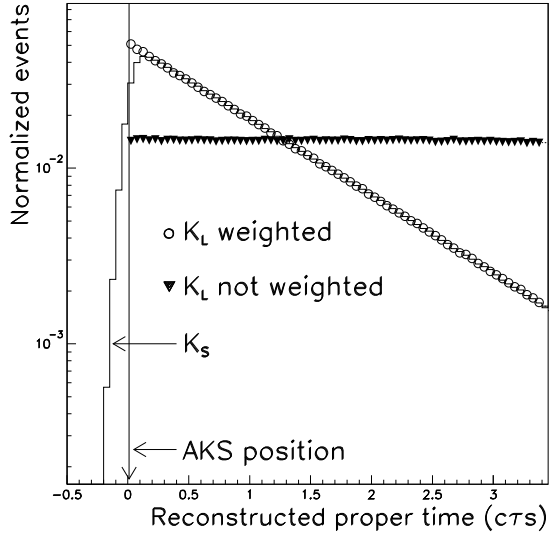,height=8cm,width=8cm}}
\end{center}
\rm
\caption{Reconstructed lifetime distribution for \kl\ $\rightarrow$ \pipic\  events before
and after weighting.}
\label{weigspec}
\end{figure}

For the kaon system decay parameters 
\taus, \taul, $\Delta M$, $\eta$ and $\phi$, 
the world averages given in the Particle Data Book ~\cite{pdb} were used. 
The \kz-\kbar\ production asymmetry factor \Dp\ is obtained by a fit to 
the data, in an energy range $E_{K}\ge$ 120~GeV where the interference term 
is significant.
After weighting, the \kl\ and \ks\ decay distributions become nearly 
identical, as shown in Fig.~\ref{weigspec}. The few \ks\ events reconstructed
upstream of the AKS are due to resolution while in the \kl\ case  the applied
cut at $\taus$=0 defines the beginning of fiducial region.
A small difference in  acceptances remains, 
related to the difference in \ks\ and \kl\ beam sizes and directions. 
This residual correction is computed using a large-statistics Monte Carlo simulation.

By adopting the proper-time weighting method, 
the \kl\ statistics effectively decrease and,  
within the fiducial region of 3.5\taus, 
it results in a \about 35\% increase in the statistical error of \R.

\subsection{Simulation tools}
The Monte Carlo precisely simulates the \kz\ beams and detector apertures. 
The simulation of the \kl\ beam includes the \ks\ component and the
interference term, using the measured production asymmetry factor \Dp. 
Particle interactions in the detector material are parametrised. 
The effect of kaon scattering and neutron interaction in the 
\ks\ collimator and the \ks\ anti-counter 
are included in the Monte Carlo 
by parametrising an energy dependent \ks\ beam halo, 
which reproduces well the  experimental distributions 
of the centre of gravity for \pipic\ and \pipin\ modes.
The measured wire inefficiencies are introduced before the event
 reconstruction. To describe the photon response in the calorimeter,
a library containing 1.7$\E{5}$ showers is used, 
generated using GEANT in 36 energy bins from 2--101~GeV 
in order to guarantee good resolution and linearity and in
 a large 51$\times$51 cell area to allow realistic energy sharing
for close particles. A parametrisation of the non-Gaussian tails is 
added to the photon energy, together with the measured electronic noise 
and known contributions from non-uniform response. 
Two other libraries are used: 
one for electrons to describe conversions or Dalitz decays, 
and another for charged pions.

To obtain an accurate acceptance correction, 
3$\E{8}$ kaon decays per mode were generated. 
After all cuts, statistics equivalent to five times the data were obtained. 
The generation is sub-divided into run periods 
that follow the spectrometer magnetic field direction, 
the beam position in the detectors, 
dead cells in the calorimeter and wire problems, 
as they are observed in the data.
Simulated events undergo standard reconstruction and analysis criteria. 

\subsection{Correction on the double ratio and systematic checks}
\begin{figure}[h!]
\begin{center}
\vspace{-.8cm}
\mbox{\epsfig{file=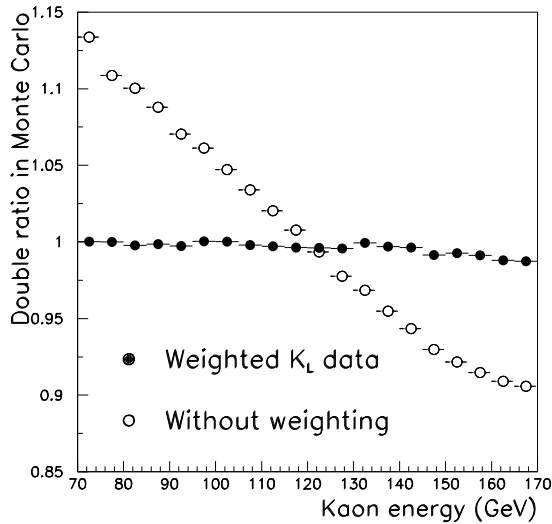,height=8cm,width=8cm}}
\end{center}
\rm
\caption{Acceptance effect to the double ratio \R\, as computed
from Monte Carlo, in kaon energy bins.
The case without weighting \kl\ data is also shown.}
\label{accepcor}
\end{figure}

\begin{figure}[h!]
\begin{center}
\vspace{-.5cm}
\mbox{\epsfig{file=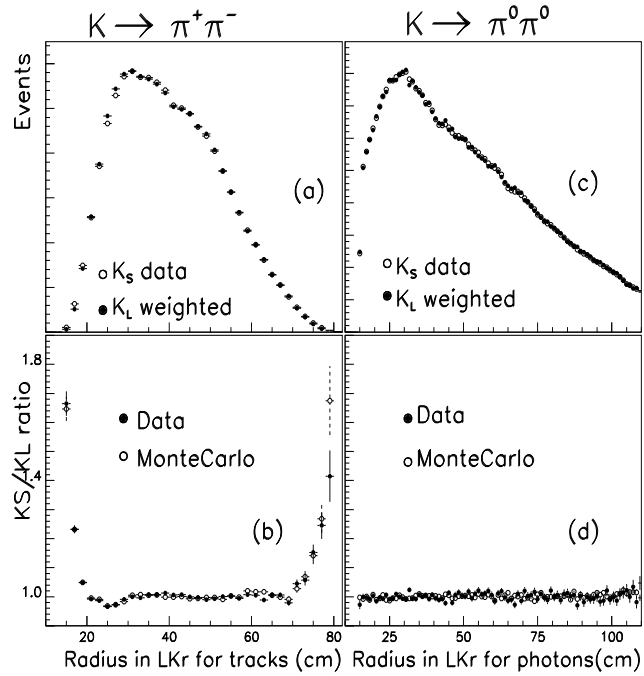,height=9.5cm,width=8.5cm}}
\end{center}
\rm
\caption{The calorimeter illumination by \pipic\ tracks 
for \ks\ and \kl\ weighted events is shown in (a), 
while the \ks/\kl\ ratio for data and MC is shown in (b). 
Similarly, the calorimeter illumination  for photons as a function of the 
radius for \pipin\ is shown in (c), 
with  the \ks/\kl\ ratio for data and MC shown in (d). Distributions are
shown for events in the kaon energy range 95-100 GeV.} 
\label{accep}
\end{figure}

The effect of the acceptance on the double ratio, as obtained from
Monte Carlo, is shown in Fig.~\ref{accepcor} in 
5~GeV kaon energy bins. It is used to derive the acceptance correction.
 For comparison, the acceptance effect without 
applying  proper time weighting to the \kl\ events is also shown.\\
The largest contribution to the correction comes from 
the difference between the \ks\ and \kl\ beams 
around the beam axis in the spectrometer for \pipic\ decays. The acceptance 
correction related to the \pipin\ mode is small.
The illumination of the calorimeter for \pipic\ and \pipin\ decays is
 shown in Fig.~\ref{accep}, in a comparison for \ks\ and weighted \kl\ events. 
The \ks/\kl\ ratio varies in the \pipic\ mode for track positions close
to the beam tube (Fig.~\ref{accep}.b). This is reproduced by the simulation, 
so the effect is absorbed into the acceptance correction. 
In the \pipin\ mode, the photon distribution in the calorimeter is 
almost identical
for \ks\ and weighted \kl\ events, as shown in Fig.~\ref{accep}.d.

The systematic uncertainties are evaluated 
by modifying certain parameters until the agreement between
experimental and simulated distributions is affected. 
Varying the \ks\ beam halo, the \kz-\kbar\ production asymmetry, 
the beam shapes in the detectors 
and the wire inefficiencies, 
the correction is stable to within  $3.3\E{-4}$.  
A systematic cross-check was done by sending 
the generated kinematic values of all 3$\E{8}$ events per mode
 of the Monte Carlo 
through a GEANT based Monte Carlo in order to identify 
biases associated to the fast simulation. 
Within the statistical accuracy, no difference was observed in the neutral mode.
In the charged mode, an event-by-event comparison 
showed a systematic difference of $(-4.6 \pm 2.3)\E{-4}$. 

While the effect is small, the average of the two evaluations,
with a systematic error increased by 2.3$\E{-4}$,
is conservatively quoted as an acceptance correction.
The final correction to \R\ for
the acceptance is: 
\begin{eqnarray*}
\DR(\mbox{acceptance}) = (+26.7 \pm 4.1(\stat) \pm 4.0(\syst))\E{-4}
\end{eqnarray*}

\section{Effects related to beam activity and detector response variations}

The identification of good \kpipi\ decays might be affected 
by the  presence of extra activity from various sources.
Additional particles can either be produced simultaneously with the
kaon candidate in the target (called `in-time'), 
or they can come from uncorrelated kaon decays (called `accidentals'). 
Accidental particles come mainly from the intense \kl\ beam, 
and the neutron and muon components accompanying it. 
In-time background is effectively suppressed for \kl\ decays given
the beam collimation and the extensive shielding, while in the 
\ks\ beam a possible additional in-time component needs to be considered.
Beam intensity variations do not affect the double ratio, 
thanks to the concurrent recording of all four modes. 
However, slow variations of the \ks/\kl\ intensity during the run 
that accidentally coincide with unaccounted variations in the charged-to-neutral event ratio 
(e.g. from a change in the detector response), 
would lead to a bias. 
The measured variance of the \ks/\kl\ intensity ratio is \about 9\%, 
while the \pipic/\pipin\ ratio, corrected for trigger inefficiency,
varied by 1--2\% during the 1998-1999 data period.
The potentially resulting effect is taken into account in the analysis 
by weighting the \ks\ events by a factor extracted from 
\ks/\kl\ intensity variation monitoring. 
The effect of this weighting on \R\ is small ($\le 3\E{-4}$),
indicating that both the beams and the detectors were
sufficiently stable during the two years of data taking. 

\subsection{Accidental activity}
The pile-up  of extra particles with a good event
may result in an event loss, 
depending on the time and space separation in the detector. 
Owing  to the simultaneous data taking, \ks\ and \kl\ events are equally  sensitive
to the accidental beam activity, to first order.
A residual effect on \R\ can be written as:
\begin{equation}
\DR = \DRint + \DRgeom
\label{eqn:twoterms}
\end{equation}
The two terms have the following origins: 
\begin{enumerate}
\item   {\bf The intensity term}, \DRint, appears 
        if there is a decorrelation between the two beams.
Describing the \ks\ and \kl\ beam intensity time dependences as I$_{S,L}$(t)=
$\langle$I$_{S,L}\rangle \times$(1+\etasl(t)), this would mean that 
$\eta_{S}$(t) and $\eta_{L}$(t) differ.  
        If in addition the mean \pipic\ and \pipin\ losses 
        (\lamc\ and \lamn\ respectively)
        are not equal an effect on the double ratio can be expected.
        Assuming that losses depend linearly on the beam intensity 
        this effect is given by:
        \begin{equation}
        \DRint =(\lamc-\lamn) \times \langle \etal(\etal-\etas) \rangle
        \label{eqn:Int-term}
        \end{equation}
\item   {\bf The illumination term}, \DRgeom, is non-zero if
        accidental losses depend on the detector illumination, 
        since the latter is slightly different for the \ks\ and \kl\ beams.
        This can influence the double ratio if it is not equal 
        for \pipin\ and \pipic\ events. The corresponding effect can
        be expressed as:
        \begin{equation}
        \DRgeom = (\delta\lamc-\delta\lamn) \times ( 1 + \langle \etal\etal \rangle)
        \end{equation}
        where $\delta\lambda$ represents the difference in losses between the \ks\ and \kl\ beams. 
\end{enumerate}  
The measurement of event losses induced by extra activity 
is necessary for evaluating these terms. 
This is done through the overlay method. 

\subsubsection{Overlay method}
\label{sec:overlay}
Losses from  accidental activity can be simulated  
by overlaying data with special beam monitor (BM) triggers.  
Beam monitors (see section~\ref{sec:othtri}) trigger 
in proportion to the \ks\ and \kl\ beam intensities, 
so that the BM triggers give a picture of the ambient activity 
and noise measured by the detectors at a given time. 
Detector information for the event 
and the BM trigger are superimposed, 
and the so-called overlaid event undergoes the usual reconstruction. 
For this purpose only BM triggers without DCH overflow condition
are used. After applying the standard analysis procedure, 
the probability of losing or gaining an event is computed
by comparing, event-by-event, the original and overlaid samples. 
This method allows the effect of the extra activity and noise
that was present when the data was recorded
to be reproduced and measured.
However, the method depends on how well 
the BM trigger content describes the intensity variations seen by the data. 
A comparative study of the extra activity in BM triggers and 
kaon events showed an agreement that was better than 97\%.       

To overcome the problem of doubling the detector noise in the overlaid event, 
leading to an overestimation of the losses, 
an alternative method is used in parallel: 
the overlaid Monte Carlo (OMC).
This consists of using the BM triggers to overlay 
noiseless Monte Carlo events, instead of overlaying data. 
A small fraction of Monte Carlo events were overlaid consecutively
with two different BM triggers 
(doubly overlaid Monte Carlo, or DOMC), 
in order to allow a direct comparison with the losses computed from overlaid data.

With these methods, 
two sources of losses and gains for good events
were identified:
\begin{itemize}
\item   The mortality due to a BM trigger. 
        A BM trigger containing extra activity 
        has an enhanced probability of killing an event in the overlay procedure. 
        The extra activity depends on the beam intensity. 
        Because of the different selection criteria, 
        this would affect the \pipic\ and \pipin\ modes asymmetrically. 
        Examples of cuts sensitive to the intensity include
        the muon veto in \pipic\ and the rejection of extra photon clusters in
\pipin.
\item   Noise in a detector.  
        An event can be lost or gained after the overlay 
        owing to migration across cut boundaries. 
        Certain selection criteria are sensitive to noise, 
        like the \chisq\ distribution in \pipin. 
        Losses caused by noise do not depend on the beam intensity.
\end{itemize}  
The overall effect of the overlay is dominated by losses.
The net effect of the overlay (losses minus gains) 
is shown in Tab.~\ref{losses-gains}.
OMC describes the reconstruction losses due to accidental activity and noise.
Losses in the charged mode are dominated by the DCH multiplicity overflow condition,  
while in the neutral mode the \chisq\ cut accounts for the bulk of the losses. 
Results obtained by comparing  DOMC and OMC  
are directly comparable to those from the overlaid data.
The agreement is reasonable in the \pipin\ mode. 
DOMC gives slightly fewer losses in the \pipic\ mode. 
This is associated with the smaller DCH multiplicity condition rate in DOMC.
\begin{table}[h!]
\begin{center}
\caption{Net effect of accidental activity, computed using
data and Monte Carlo events overlaid with BM triggers.The LKr noise effect is subtracted. The DOMC column 
represents differences between first and second overlay.}
\label{losses-gains}
\vspace{2mm}
\begin{tabular}{|l|c|c|c|c|}
\hline
{\bf{Losses--Gains}}            &  OMC          & DOMC          & Data overlay  \\ \hline
{\boldmath \piopio}             & 0.8\%         & 1.5\%         & 1.4\%         \\ \hline
{\boldmath \pipic }             & 1.8\%         & 2.5\%         & 2.8\%         \\ \hline
{\boldmath $\pipic-\piopio$}    & 1.0\%         & 1.1\%         & 1.4\%         \\ \hline
\end{tabular}
\end{center}
\end{table}

\begin{figure}[h!]
\begin{center}
\vspace{-.5cm}

\mbox{\epsfig{file=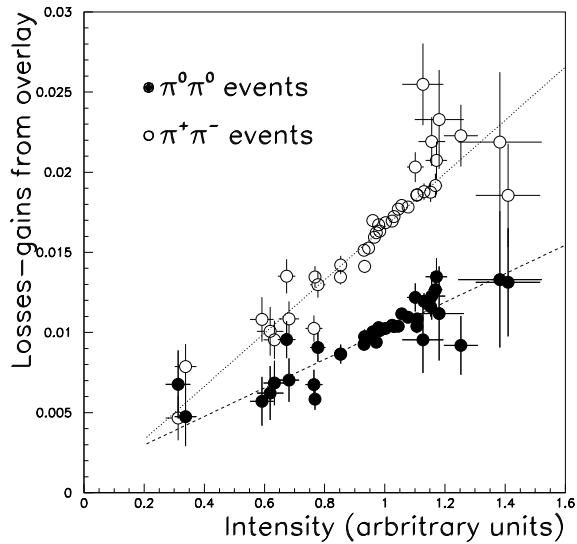,height=8cm,width=8cm}}
\end{center}
\rm
\caption{Losses as a function of the intensity seen by the BM trigger, computed
by the overlay method. Lines are plotted to guide the eye.} 
\label{loslin}
\end{figure}

In order to explore the losses 
throughout the entire range of beam intensity, 
information from the beam monitor, 
placed at the end of the beam tube, is used. 
In Fig.~\ref{loslin}, the net accidental effect,  
computed from OMC, is shown for \pipic\ and \pipin\ events 
as a function of the recorded intensity 
at the time of the BM trigger used for the overlay. 
Losses increase faster in \pipic\  than in  \pipin\ events. 
The overall behaviour is compatible with the assumption 
that the dependence of accidental losses with the rate is linear.   

In what follows, the conservative value computed from data overlay 
is assigned to the difference of the effect of accidental activity on \pipic\ and 
\pipin\ events,
namely $\lamc-\lamn = (1.4 \pm 0.7)\%$.

\subsubsection{Effect on the Double Ratio}
Having measured the accidental losses for \pipic\ and \pipin\ events, 
the two terms of Eq.~\ref{eqn:twoterms} can be computed.
The intensity variation term, \DRint, depends upon 
how well the variations of the two beams ($\eta_{S},\eta_{L}$) 
mimic one other, 
as well as upon the differential losses between the \pipic\ and \pipin\ 
modes ($\lambda^{+-},\lambda^{00}$), 
according to Eq.~\ref{eqn:Int-term}. 
Several estimators have been used to study a possible 
difference in the activity of the two beams, 
based on accidental information in good events.
Among these are the number of hits in the AKL counters, 
the number of accidental clusters in the LKr calorimeter, 
and the number of extra tracks,
all of which have been
measured in both \ks\ and \kl\ events. 
The corresponding occurrence rates for these estimators are similar in the
two beams. An independent method to measure the difference 
involves examining the correlation between the \ks\ and \kl\ beams.  
All results agree that, 
to within 1\%, 
there is no intensity difference seen by \ks\ and \kl\ events. 
So from Eq.~\ref{eqn:Int-term} we obtain  $\DRint =\pm 3\E{-4}$, 
where an additional, conservative factor of two is used in the uncertainty, 
to account for a possible non-linear dependence of losses with the intensity.

The illumination term, \DRgeom, is estimated by the 
overlay method with data or Monte Carlo events.
Both computations agree that there is no significant
effect within available precision. 
The interpretation of the
result is that  all sources of losses are uniformly
distributed in the acceptance. 
An upper bound on the illumination term can be quoted, 
according to the statistical significance of the measurement, namely $\DRgeom = 
\pm 3\E{-4}$.
Finally, the uncertainty on the double ratio from accidental
activity is:
\begin{eqnarray*}
\DR = \DRint + \DRgeom = \pm 4.2\E{-4}
\end{eqnarray*}

\subsection{In-time activity}
Particles produced in the \kl\ target, in time with the \kl\, 
are well suppressed by the 120 m of collimation and magnets. 
In the \ks\ case, the 6~m collimator occasionally allows in-time particles 
to survive into the decay region, 
and to be detected together with a good kaon decay.

To put bounds on such effects, 
the \piopio\ sample accumulated in single \ks\ runs is used. 
The low intensity of the \ks\ beam ensures the absence of 
accidental particles. The \piopio\ mode was chosen for the clean 
identification of the number of in-time clusters; in the charged mode, 
the extended and irregular hadronic showers could bias the study.
By removing the extra (in-time) cluster cut, 
the \pipin\ signal is enriched with the following events:
\pipin\ events which have a photon conversion, giving 5-cluster final modes
and \pipin\  events with additional in-time particles. 
Only in the case of photon conversion is the reconstructed \chisq\ deteriorated. 
By comparing the experimental \chisq\ distribution of \twopio\ events with five clusters 
to the one predicted by Monte Carlo solely for conversions, 
an excess of $(0.7 \pm 0.4)\E{-4}$ was found.  
This would correspond to the in-time activity,  
and has been cross-checked using various time windows for the extra-activity.
Therefore, an upper bound on the in-time contamination in the \ks\ beam can be 
set to 1$\times 10^{-4}$. This is included in the uncertainty on accidental
activity in Tab.~\ref{tab:syst}.

\section{Result}
\label{sec:result}

\begin{figure}[h!]
\vspace{-1cm}
 \begin{center}
  \includegraphics[width=0.49\textwidth]{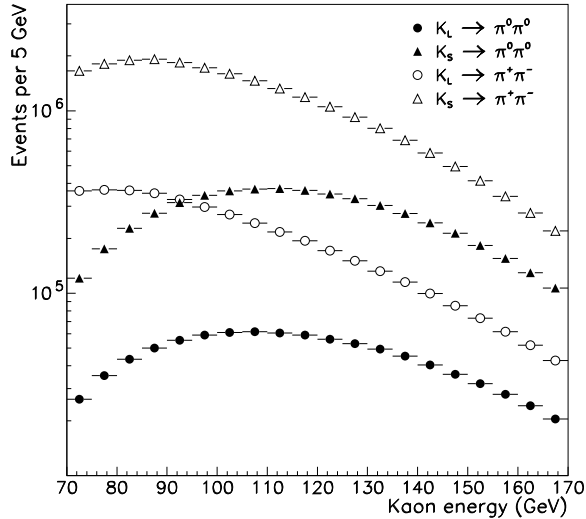}
  \caption{Energy distribution of selected events after accounting for
    mistagging and after proper-time weighting of \KL\ events.}
  \label{fig:spect}   
 \end{center}
\end{figure}

The number of events passing all selection criteria, 
taking into account \ks\ mistagging, 
is shown in Tab.~\ref{tab:stat} and their energy distribution in
Fig.~\ref{fig:spect}. 
The final result is computed by dividing the data into 20 bins of kaon
energy from 70 to 170 GeV, 
and calculating  the double ratio for each bin using proper-time weighted \kl\ samples. 
All corrections are applied to each bin separately, 
and the results are averaged.
In order to avoid a statistical bias, 
instead of the simple weighted mean, 
the bins are averaged using a statistically unbiased logarithmic estimator. 
The average effects of the corrections 
on the raw double ratio, together with their uncertainties, 
are summarised in Tab.~\ref{tab:syst}.

\begin{table}
\begin{center}
\caption{Number of selected events after accounting for mistagging.}
\label{tab:stat}
\begin{tabular}{|l|r|r|r|}
\hline
&\multicolumn{3}{c|}{Statistics in millions} \\ \hline
& 1998 & 1999 & total \\ \hline
$\KL \rightarrow \pipin$ & 1.047 & 2.243 & 3.290 \\
$\KS \rightarrow \pipin$ & 1.638 & 3.571 & 5.209 \\
$\KL \rightarrow \pipic$ & 4.541 & 9.912 & 14.453 \\
$\KS \rightarrow \pipic$ & 6.910 & 15.311 & 22.221 \\ \hline
Statistical error & 18.0$\times$10$^{-4}$ &  12.2$\times$10$^{-4}$ & 
                  10.1$\times$10$^{-4}$\\
\hline
\end{tabular}
\end{center}
\end{table}

\newlength{\acclen} \settowidth{\acclen}{$\pm$ 4.0}

\begin{table}[htb]
\begin{center}
\caption{Corrections and systematic uncertainties on the double
ratio. In order to obtain the 
 effect on \Ree, the numbers must be divided by a factor of 6
 (Eq. \ref{eq:doub}).} 
\label{tab:syst}
\begin{tabular}{|l|rr|} \hline 
&\multicolumn{2}{c|}{in 10$^{-4}$} \\ \hline
\pipic\ trigger inefficiency                    &  $-3.6$       & $\pm$ 5.2  \\
AKS inefficiency                                &  $+1.1$       & $\pm$ 0.4  \\
Reconstruction \begin{tabular}{@{}l} of \pipin\ \\ of \pipic\ \end{tabular} &
\begin{tabular}{r@{}}    ---   \\  $+2.0$  \end{tabular} &
\begin{tabular}{r@{}}   $\pm$ 5.8 \\  $\pm$ 2.8  \end{tabular} \\
Background \begin{tabular}{@{}l} to \pipin\ \\ to \pipic\ \end{tabular} &  
\begin{tabular}{r@{}} $-5.9$ \\ $+16.9$ \end{tabular} &
\begin{tabular}{r@{}} $\pm$ 2.0  \\ $\pm$ 3.0  \end{tabular} \\
Beam scattering                                 &  $-9.6$       & $\pm$ 2.0  \\
Accidental tagging                              &  $+8.3$       & $\pm$ 3.4  \\
Tagging inefficiency                            &  ---          & $\pm$ 3.0  \\
Acceptance \begin{tabular}{@{}l} statistical \\ systematic \end{tabular}   
                              & $+26.7$ & 
     \begin{tabular}{r@{}} $\pm$ 4.1 \\ $\pm$ 4.0 \end{tabular}  \\
Accidental activity                             &  ---          & $\pm$ 4.4  \\ 
Long term variations of \KS/\KL\  &  ---   & $\pm$ 0.6  \\ \hline 
Total                                           & $+35.9$       & $\pm$ 12.6 \\ \hline 
\end{tabular}
\end{center}
\end{table}

\begin{figure}[h!]
\vspace{-1cm}
 \begin{center}
  \includegraphics[width=0.49\textwidth]{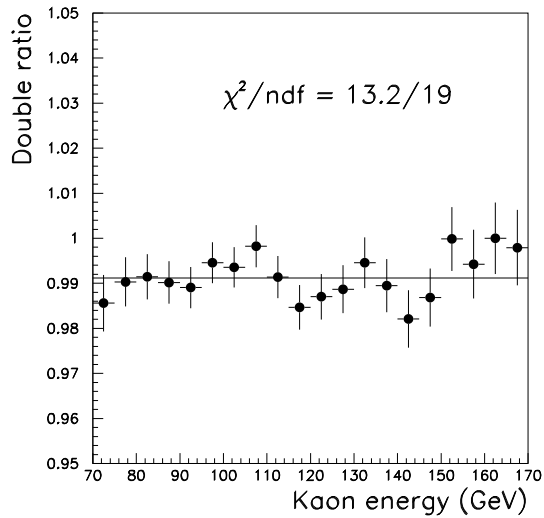}
  \caption{Corrected double ratio as a function of
    the kaon energy.}
  \label{fig:r_e}   
 \end{center}
\end{figure}

\begin{figure}[h!]
 \begin{center}
\vspace{-1cm}
  \includegraphics[width=0.51\textwidth]{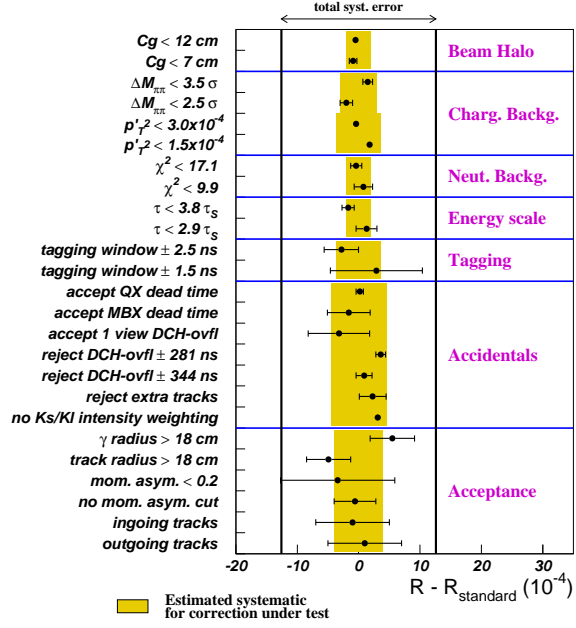}
  \caption{Stability of the double ratio with variations of the
    selection cuts. The grey band shows the uncertainty related to the
    cut concerned. \R$_{\mathrm{standard}}$ represents the value from
    Eq.~\ref{eq:rstand}.}
  \label{fig:systvar_general}   
 \end{center}
\end{figure}

After applying all corrections to the double ratio, the result is
\begin{equation}
\label{eq:rstand}
R = (0.99098 \pm 0.00101(\stat) \pm 0.00126(\syst))
\end{equation}
The stability of the result as a function of various measurement
parameters was studied extensively. 
The double ratio as a function of the kaon energy is shown in Fig.~\ref{fig:r_e}.
The size of the assigned systematic uncertainties 
was tested by varying the most important selection cuts.
By changing the background-rejection cuts, 
even large variations of the amount of subtracted background 
did not cause any shift in the result greater than 
the estimated uncertainty on the background. 
Similarly, no excessive deviations of the measurement were observed by varying
cuts related to the acceptance, 
\ks-proton tagging, 
uncorrelated beam activity, 
the decay volume and beam scattering (see Fig.~\ref{fig:systvar_general}). 

The stability of the result was also tested with respect to 
the time variation of the measurement conditions. 
In this context,
the double ratio was measured in run periods defined by technical accelerator stops. 
Further checks were performed as functions of spill time, 
SPS revolution phase, 
50~Hz mains phase, 
the spectrometer magnet polarity setting, 
and the time of day. 

Another test consisted of splitting
the \pipic\ data sample into converging and diverging topologies,
depending on the orientation of the track-charges in the magnetic field
of the spectrometer, thereby selecting different acceptances and momentum resolutions. 
The steadiness of the drift-chamber read-out overflow rejection was tested by
accepting a special class of overflow-events which affect only one
pair of staggered drift chamber planes, 
and have only moderate effects
on the trigger and reconstruction efficiencies.
The variations observed in all tests are well within the assigned
systematic uncertainties.

The data filtering,
selection and calculation of the corrections were performed by
several independent groups. 
One analysis adopting a different scheme of data
compacting and filtering was performed in parallel to the one presented.
In this analysis, several selection criteria were chosen differently, 
and procedures (such as the background subtraction) 
were conceived independently. 
This analysis obtained a result fully confirming the above measurement.

Converting the corrected double ratio using Eq.~\ref{eq:doub}, 
the following 
measurement of the direct CP violation parameter \Ree\ is obtained:
\begin{equation}
\Ree = (15.0 \pm 1.7(\stat) \pm 2.1(\syst)) \E{-4}
\end{equation}
Adding the two uncertainties in quadrature gives:
\begin{equation}
\Ree = (15.0 \pm 2.7) \E{-4}
\end{equation}
The combined NA48 \Ree\ measurement, including the 
result from the 1997 data~\cite{res2}, is:
\begin{equation}
\Ree = (15.3 \pm 2.6) \E{-4}
\end{equation}
where correlated uncertainties are taken into account.

\section{Conclusions}

The NA48 measurement of the direct CP violation parameter \Ree, 
based on the analysis of data from 
the 1998 and 1999 data taking periods, 
is in good agreement with the result from the 1997 data. 
The combination of both
results deviates from zero by 5.9 standard deviations.

\section{Acknowledgements}

It is a pleasure to thank the technical staff of the participating
laboratories and universities for their efforts in the design and
construction of the apparatus, in the operation of the experiment, and
in the processing of the data.

\end{document}